\documentclass[aps,10pt,prd,groupedaddress,nofootinbib,notitlepage,eqsecnum,preprintnumbers]{revtex4-1}

\usepackage[utf8]{inputenc}
\usepackage{hyperref}
\usepackage{xcolor}
\usepackage{graphicx}
\usepackage{amsmath,amssymb,amsbsy,amstext,amsthm}
\usepackage{mathtools}
\usepackage{amsfonts}
\usepackage{bm}
\usepackage{tabularx}
\usepackage{array}

\newcolumntype{C}{>{\collectcell\docellC}c<{\endcollectcell}}
\newcolumntype{L}{>{\collectcell\docellL}c<{\endcollectcell}}
\newcolumntype{R}{>{\collectcell\docellR}c<{\endcollectcell}}

\makeatletter
\providecommand{\@nameedef}[1]{\expandafter\edef\csname#1\endcsname}
\newcommand{\docell}[2]{%
  \sbox\equalizedtablebox{#2}%
  \ifdim\wd\equalizedtablebox>\@nameuse{finallen\theequalizedtable}\relax
    \global\@nameedef{finallen\theequalizedtable}{\the\wd\equalizedtablebox}%
  \fi
  \makebox[\@nameuse{startinglen\theequalizedtable}][#1]{#2}%
}
\newcommand{\docellC}[1]{\docell{c}{#1}}
\newcommand{\docellL}[1]{\docell{l}{#1}}
\newcommand{\docellR}[1]{\docell{r}{#1}}
\newcounter{equalizedtable}
\newsavebox\equalizedtablebox

\makeatother

\begin{document}
\preprint{YITP-20-156}
\title{Gravitational wave constraints on the primordial black
hole dominated early universe}

\author{\textsc{Guillem Dom\`enech$^{a}$}}
    \email{{domenech}@{pd.infn.it}}
\author{\textsc{Chunshan Lin$^{b}$} }
    \email{{chunshan.lin}@{uj.edu.pl}} 
\author{\textsc{Misao Sasaki$^{c,d,e}$} }
    \email{{misao.sasaki}@{ipmu.jp}}

\affiliation{$^{a}$ \small{INFN Sezione di Padova, I-35131 Padova, Italy}\\
      $^{b}$\small{Faculty of Physics, Astronomy and Applied Computer Science, Jagiellonian University, 30-348 Krakow, Poland}\\
      $^{c}$\small{Kavli Institute for the Physics and Mathematics of the Universe (WPI), Chiba 277-8583, Japan}\\
      $^{d}$\small{Center for Gravitational Physics, Yukawa Institute for Theoretical Physics, Kyoto University, Kyoto 606-8502, Japan}\\
      $^{e}$\small{Leung Center for Cosmology and Particle Astrophysics, National Taiwan University, 
      Taipei 10617, Taiwan}
    }

\begin{abstract}
We calculate the gravitational waves (GWs) induced by the density fluctuations due to inhomogeneous distribution of primordial black holes (PBHs)
 in the case where PBHs eventually dominate and reheat the universe by Hawking evaporation.
 The initial PBH density fluctuations are isocurvature in nature. 
 We find that most of the induced GWs are generated right after evaporation, 
 when the universe transits from the PBH dominated era to the radiation dominated era and the curvature perturbation starts to oscillate wildly.
 The strongest constraint on the amount of the produced GWs comes from the big bang nucleosynthesis (BBN).
 We improve previous constraints on the PBH fraction and find that it cannot exceed $10^{-3}$. Furthermore, 
 this maximum fraction decreases as the mass increases and reaches $10^{-9}$ for 
 $M_{\rm PBH}\sim 5\times10^8 {\rm g}$, which is the largest mass allowed by the BBN constraint
 on the reheating temperature.  Considering that PBH may cluster above a given clustering scale, we also derive a lower bound 
 on the scale of clustering. Interestingly, the GW spectrum for $M_{\rm PBH}\sim 10^4 -10^8 {\rm g}$ enters the observational 
 window of LIGO and DECIGO and could be tested in the future. Although we focus on the PBH dominated early universe in this paper,
  our methodology is applicable to any model with early isocurvature perturbation.
\end{abstract}
 \maketitle

\section{Introduction}

Primordial black holes (PBHs) form upon the collapse of large primordial fluctuations after horizon reentry \cite{Zeldovich:1967lct,Hawking:1971ei,Carr:1974nx,Meszaros:1974tb,Carr:1975qj,Khlopov:1985jw}. 
Their lifetime under Hawking evaporation is larger than the age of the universe if their masses are greater than $10^{15}{\rm g}$. 
This kind of long-lived PBHs are attracting a lot of attention as potential candidates for explaining dark matter \cite{Carr:2016drx,Inomata:2017okj}, 
the binary black holes responsible for the LIGO/VIRGO gravitational waves (GW) detection \cite{Bird:2016dcv,Clesse:2016vqa,Sasaki:2016jop}, 
the seeds of supermassive black holes \cite{Kawasaki:2012kn,Carr:2018rid}, the microlensing events by planet-mass objects found by OGLE \cite{2017Natur.548..183M,Niikura:2019kqi} and the planet 9 possibly responsible for the unusual orbit of trans-Neptunian objects \cite{Scholtz:2019csj,Witten:2020ifl}. See Ref.~\cite{Sasaki:2018dmp} for an extensive review. 
Furthermore, PBH could even explain the baryon asymmetry of the universe \cite{Baumann:2007yr,Hook:2014mla,Hamada:2016jnq,Carr:2019hud}. Conversely, PBH lighter than $10^{15}{\rm g}$ provide an interesting mechanism to reheat the universe by Hawking evaporation \cite{Lidsey:2001nj,Hidalgo:2011fj}. 

Irrespective of the mass, PBH formation is inevitably accompanied by an abundant production of secondary GWs. 
Right after horizon reentry, large primordial fluctuations oscillate with an amplitude large enough to source, at second order in perturbation theory, detectable GW \cite{tomita,Matarrese:1992rp,Matarrese:1993zf,Matarrese:1997ay,Noh:2004bc,Carbone:2004iv,Nakamura:2004rm,Ananda:2006af,Baumann:2007zm,Osano:2006ew}. With newly derived analytical formulas \cite{Espinosa:2018eve,Kohri:2018awv,Domenech:2019quo,Domenech:2020kqm}, the so-called induced GW are also drawing a lot of interest since they could lie within the future observational window of LISA \cite{Audley:2017drz}, Taiji \cite{Guo:2018npi}, Tianqin \cite{Luo:2015ght}, DECIGO \cite{Seto:2001qf,Yagi:2011wg}, AION/MAGIS \cite{Badurina:2019hst}, ET \cite{Maggiore:2019uih} and PTA \cite{Lentati:2015qwp,Bian:2020bps}. Not only they are crucial PBH counterparts \cite{Espinosa:2018eve,Cai:2018dig,Bartolo:2018rku,Yuan:2019udt} but constitute powerful probes of the primordial curvature power spectrum on small scales \cite{Inomata:2018epa,Gow:2020bzo} and the thermal history of the universe \cite{Cai:2019cdl,Hajkarim:2019nbx,Domenech:2019quo,Domenech:2020kqm}.

It has been recently noted in an interesting paper \cite{Papanikolaou:2020qtd} that primordial fluctuations in the curvature perturbation might not be the only source of induced GWs. In fact, after formation, PBHs are randomly distributed in space according to Poisson statistics to a good approximation. Thus, although on average the PBH gas behaves as a homogeneous pressure-less matter, or for short dust, the spatially inhomogeneous distribution of PBH gives rise to density fluctuations. Interestingly, the requirement that the GWs induced during the PBH dominated era by such PBH density fluctuations do not backreact onto the background leads to a substantial upper bound on the fraction of PBH at formation \cite{Papanikolaou:2020qtd}. We extend the work of Ref.~\cite{Papanikolaou:2020qtd} and study the GWs induced during the PBH dominated era in more detail, paying special attention to the GWs generated right after evaporation. In particular, we find $(i)$ a stronger constraint on the initial PBH fraction and $(ii)$ a GW spectrum induced by PBH density fluctuations which might be observable by future GWs detectors, depending on the initial PBH mass.

As commonly assumed, we consider that PBHs formed during an epoch of early radiation domination (eRD). The resulting PBH density fluctuations can essentially be regarded as early isocurvature fluctuations \cite{Papanikolaou:2020qtd}.  The isocurvature nature of the PBH density fluctuations is clear in the following simplified picture. First, assume that PBH form simultaneously in a perfectly homogeneous universe. If we treat PBHs as a dust fluid, PBH formation may be regarded as a transition of a fraction of radiation into dust matter. While the total energy density remains homogeneous at the transition, PBHs are distributed randomly and, therefore, the energy density of the dust fluid is inhomogeneous. This inhomogeneity is isocurvature because that the total energy density is homogeneous.
At a later time, if the initial fraction of PBH is large enough, PBH eventually overcome the energy density of the radiation component and dominate the universe. In the transition from the eRD to an early matter domination (eMD), the early isocurvature component is converted to curvature perturbation. When this occurs, the resulting curvature perturbation turns into a source for secondary GWs. Since the PBH density fluctuations are significant on scales comparable to the mean separation of PBH at the time of formation \cite{Papanikolaou:2020qtd}, one expects a sizable amount of induced GW at this scale. Additionally, if PBH cluster below a given clustering scale, then the spectrum of fluctuations is even more enhanced on the smallest scales. The case of the induced GWs generated by the early adiabatic perturbation in the PBH reheating scenario is very well explained in Ref.~\cite{Inomata:2020lmk}. 

Induced GW generated during an eMD epoch have been studied in detail in Refs.~\cite{Inomata:2019zqy,Inomata:2019ivs} 
and have their peculiarities. Since the curvature perturbation is constant in time on all scales during eMD, the source term for the secondary GWs is constant in time as well. 
What might seem as a gauge artifact, as probably a constant term might be gauged away, is converted to a large amount of 
induced GWs at the transition from eMD to the last radiation domination (lRD). 
If the transition is sudden, the production of GWs is abundantly enhanced. 
Following Ref.~\cite{Inomata:2019ivs}, this can be understood as follows. 
The density fluctuations during the eMD grow proportionally to the scale factor. 
In a sudden reheating, the matter fluid is vaporized into radiation and its fluctuations, 
which developed an enhanced amplitude, start to oscillate wildly. 
The resulting enhanced induced GW spectrum is much larger than that generated during the eMD. 
Such is the case of the PBH reheating scenario, where the final evaporation is almost instantaneous. Although the contribution due to the eMD to lRD transition was mentioned in Ref.~\cite{Papanikolaou:2020qtd}, they solely focused on the GW spectrum generated during eMD due to PBH domination.

In this paper, we extend the work of Ref.~\cite{Papanikolaou:2020qtd} and show that the induced GWs due to the sudden evaporation makes up the largest contribution.
We obtain a stronger upper bound on the initial fraction of PBH, demanding that the amount of thus produced GWs is 
not in conflict with big bang nucleosynthesis (BBN) constraints on the effective number of species. We also show that for a certain range of PBH masses, the GW spectrum might enter the observational window of future GWs detectors.
We note that although we focus on the case of the PBH dominated early universe, 
our formalism is general enough that it can be applied to any early isocurvature model. 

This paper is organized as follows. In Section~\ref{Sec:isocurvature} we deal with the isocurcature density fluctuations 
in the PBH reheating scenario. We first review the evolution of the curvature perturbation
 in the presence of an early isocurvature perturbation in a universe composed of radiation and matter fluids 
 in Section~\ref{subsec:isoevolution}. We then review in Section~\ref{subsec:pbhreheating} the PBH reheating scenario 
 and discuss the relevant scales. In Section~\ref{Sec:IIGWS} we compute the induced GW spectrum after the early isocurvature
 perturbation is converted to the curvature perturbation. We derive an upper bound on the initial fraction of PBHs and on
  the amount of possible clustering. We summarize our work and discuss our results in Section~\ref{sec:conclusions}. 
  In the appendices we provide several details of our calculations.

\section{Isocurvature and primordial black hole reheating \label{Sec:isocurvature}}

In this section, we review the PBH reheating scenario in which PBHs form at a given time after inflation during eRD, eventually dominate the universe and evaporate in the end. 
In this first part we follow closely Refs.~\cite{Papanikolaou:2020qtd} and \cite{Inomata:2020lmk}. Let us consider for simplicity 
that the PBH form by the collapse of a sharply peaked primordial spectrum and, therefore, the mass function of PBH
 is essentially monochromatic. Let us also assume that the fraction of PBHs at formation, defined by \cite{Sasaki:2018dmp}
\begin{align}\label{eq:beta}
\beta\equiv \frac{\rho_{{\rm PBH},f}}{3 H_f^2M_{\rm pl}^2},
\end{align}
is large enough so that PBHs eventually dominate the universe. 
In the above, $\rho_{{\rm PBH}}$ refers to the mean PBH energy density, $H$ is the Hubble parameter, 
$M_{\rm pl}$ is the reduced Planck mass, and the subscript $f$ refers to a quantity at the time of formation. 
We later present the necessary conditions on $\beta$ for a successful PBH reheating.  
The mass of the freshly formed PBH is proportional to the mass inside the horizon at the time of formation, namely \cite{Sasaki:2018dmp}
\begin{align}
M_{{\rm PBH},f}=\frac{4\pi\gamma M_{\rm pl}^2}{H_f},
\end{align}
where $\gamma\sim 0.2$ during RD \cite{Carr:1975qj}. Note that from the time of formation and on 
PBHs start to evaporate by emitting Hawking radiation \cite{Hawking:1974sw}. 
For this reason, we always include the subscript $f$ in the initial PBH mass since it decreases with time. 
After formation, PBHs are essentially evenly distributed in space following the Poisson statistics 
with a mean physical separation at formation given by \cite{Papanikolaou:2020qtd}
\begin{align}
d_f\equiv\left(\frac{3M_{{\rm PBH},f}}{4\pi\rho_{{\rm PBH},f}}\right)^{1/3}=\gamma^{1/3}\beta^{-1/3}{H_f}^{-1}\,.
\end{align}
As the PBH formation is a rare event arising from the large amplitude tail of the probability distribution function of the density perturbation,
it is natural to assume that the PBH spatial distribution follows the Poisson statistics. This means that the PBH density perturbation spectrum reads \cite{Papanikolaou:2020qtd}
\begin{align}
\langle\delta\rho_{\rm PBH}(k)\delta\rho_{\rm PBH}(k')\rangle=\frac{4\pi}{3}\left(\frac{d}{a}\right)^3\rho_{\rm PBH}^2\,\delta(k+k')\,,
\end{align}
where $k$ is the comoving wavenumber and we took into account that the mean separation redshifts with the scale factor, i.e. $d=d_f(a_f/a)$. 
The above smoothed spectrum of the density fluctuations is valid up to 
the coarse grained scale, that is, the comoving ultra-violet (UV) cut-off \cite{Papanikolaou:2020qtd}
\begin{align}
k_{UV}\equiv\frac{a}{d}=a_fH_f\beta^{1/3}\gamma^{-1/3}\,.
\end{align}

Now, assuming that in the eRD the PBH energy density is negligible compared 
to that of radiation at the time of formation, the PBH density fluctuations are isocurvature in nature. This is clear by noting that the PBH isocurvature perturbation is given by
\begin{align}\label{eq:isocurvature2}
S=\frac{\delta \rho_{\rm PBH}}{\rho_{\rm PBH}}-\frac{3}{4}\frac{\delta \rho_r}{\rho_r}
=\frac{\delta \rho_{\rm PBH}}{\rho_{\rm PBH}}+\frac{3}{4}\frac{\delta \rho_{\rm PBH}}{\rho_r}
\approx\frac{\delta \rho_{\rm PBH}}{\rho_{\rm PBH}}\quad {\rm for}~\rho_r\gg\rho_{\rm PBH}\,.
\end{align}
where the subscript $r$ refers to radiation, and the second equality follows from the isocurvature property, namely we ignore the initial density fluctuation, i.e. 
$\delta\rho_{{\rm PBH},f}+\delta\rho_{r,f}=0$.\footnote{There exists an adiabatic perturbation that gave rise to the PBH formation. 
In this paper we ignore its effect. This is justified as the adiabatic and isocurvature modes may be separately treated 
 in linear perturbation theory.}
Thus, the initial isocurvature fluctuation is given by
\begin{align}\label{eq:isocurvature}
S_f\approx\frac{\delta \rho_{{\rm PBH},f}}{\rho_{{\rm PBH},f}}\,.
\end{align}
The dimensionless initial isocurvature power spectrum then reads\footnote{The dimensionless power spectrum
   is defined by \begin{align}
    \langle S(k)S(k')\rangle=\frac{2\pi^2}{k^3} {\cal P}_{S}(k)\delta(k+k')\,.
  \end{align}
} 
\begin{align}\label{eq:PS}
{\cal P}_{S}(k)=\frac{2}{3\pi}\left(\frac{k}{k_{UV}}\right)^3\,.
\end{align}
It should be noted that the initial isocurvature perturbation will not source gravitational waves until PBHs dominate the universe, as
GWs can be generated only by fluctuations in the total energy momentum tensor (see Sec.~\ref{Sec:IIGWS} or App.~\ref{App:basicequations}). We may also generalize Eq.~\eqref{eq:PS} to include the effects of possible clustering of PBHs,
  e.g. due to a scale-dependent local non-gaussianity of the primordial curvature perturbation, as \cite{Suyama:2019cst,Young:2019osy}
\begin{align}\label{eq:PS2}
{\cal P}_{S}(k)\to P_S(k)\left(\frac{k}{k_{UV}}\right)^3\left(1+F_{\rm cluster}\left(\frac{k}{k_{\rm cluster}}\right)^n\right)\,,
\end{align}
where $n>0$ since clustering would increase the power on small scales that $k_{\rm cluster}<k<k_{UV}$, and $F_{\rm cluster}$ is a constant parameterizing the strength of clustering.

\subsection{Conversion of isocurvature into curvature perturbation\label{subsec:isoevolution}}

Let us now review the evolution of early isocurvature perturbations in a radiation-matter universe studied in Refs.~\cite{Kodama:1986fg,Kodama:1986ud,Mukhanov:2005sc}. 
We focus on an initially purely isocurvature perturbation, as the primordial curvature perturbation can be treated separately at linear order
in perturbation theory.
An isocurvature perturbation, as defined in \eqref{eq:isocurvature}, stays constant on superhorizon scales and source a curvature perturbation. 
The conversion from isocurvature to curvature perturbations is complete once the main component of the isocurvature perturbation 
dominates the universe. In order to see such a conversion, we consider a perturbed metric in the Newton gauge as
\begin{align}
ds^2=a^2(\tau)\left[-(1+2\Psi)d\tau^2+(\delta_{ij}+2\Phi\delta_{ij}+h_{ij})dx^idx^j\right]\,,
\end{align}
where $\tau$ is the conformal time, $a$ the scale factor, $\Psi$ and $\Phi$ respectively are the lapse and curvature perturbations and $h_{ij}$ is 
the transverse-traceless component of the spatial metric, i.e., it carries tensor degrees of freedom. 
At the background level, the scale factor is given in terms of the conformal time as 
\begin{align}
a(\tau)=a_{\rm eq}\left(2\xi+\xi^2\right)\quad {\rm where}\quad\xi\equiv(\sqrt{2}-1)(\tau/\tau_{\rm eq})\,,
\end{align}
and the subscript eq refers to the time of PBH-radiation equality, that is when $\rho_{\rm PBH}=\rho_r$ (not to be confused with the later two radiation-matter equalities, namely the one during the PBH evaporation era, and the other one around redshift $z\sim 3400$). For the moment we neglect any energy transfer between components and, therefore, we have that $\rho_{\rm PBH}\propto a^{-3}$ and $\rho_r\propto a^{-4}$. This yields a good approximation since Hawking radiation has a negligible effect onto the PBH mass at the initial stage of evaporation.

At first order in perturbation theory we have that the curvature perturbation $\Phi$ obeys \cite{Mukhanov:2005sc}
\begin{align}\label{eq:curvatureeom}
\Phi''+3{\cal H}(1+c_s^2)\Phi'+({\cal H}^2(1+3c_s^2)+2{\cal H}')\Phi-c_s^2\Delta\Phi=\frac{a^2\rho_{\rm PBH}}{2M_{\rm pl}^2}c_s^2S\,,
\end{align}
where ${\cal H}\equiv a'/a=aH$ is the conformal Hubble parameter, a prime denotes derivative with respect to conformal time and we defined as usual
\begin{align}
c_s^2\equiv\frac{4}{9}\frac{\rho_r}{\rho_{\rm PBH}+\frac{4}{3}\rho_r}\,,
\end{align}
which smoothly interpolates between $c_s^2\to1/3$ in the eRD to $c_s^2\to 0$ deep inside PBH domination. 
For an isocurvature perturbation, $\Phi=0$ initially. But it evolves to an adiabatic perturbation as the universe becomes dominated by matter. 
The resulting adiabatic perturbation sources GWs during the PBH dominated era and the post-evaporation era of RD.
One can check that in the case of solely MD, $\Phi=const.$ is a solution to Eq.~\eqref{eq:curvatureeom} for all scales. 
The value of $\Phi$ in eMD depends on when the mode enters horizon. 
Let us denote the mode that crosses the horizon at PBH-radiation equality by $k_{\rm eq}={\cal H}_{\rm eq}$.
For $k<k_{\rm eq}$, $\Psi$ becomes a constant after the mode enters the horizon at MD.
For  $k>k_{\rm eq}$, the mode enters the horizon during eRD, and $\Phi$ decays in proportion to $a^{-2}$  
until eMD is reached. In both cases, one finds analytical solutions that may be approximately 
 given by \cite{Kodama:1986fg,Kodama:1986ud}
\begin{align}\label{eq:conversionfinal}
\Phi_{\rm eMD}(k;a\gg a_{\rm eq})=S\left\{
\begin{aligned}
&\frac{1}{5}\qquad  & k\ll k_{\rm eq}\\
&\frac{3}{4}\left(\frac{k_{\rm eq}}{k}\right)^2\qquad & k\gg k_{\rm eq}
\end{aligned}
\right.\,.
\end{align}
Later we use Eq.~\eqref{eq:conversionfinal} as initial conditions to calculate the induced GWs generated at evaporation. 
It should be noted that solution \eqref{eq:conversionfinal} is independent on the initial amplitude, $S$, of the early isocurvature perturbation, 
and hence this transfer function can be applied to any isocurvature model with eMD.

After PBHs evaporate, the universe enters the lRD, and $\Phi$ inside the horizon starts to oscillate and decay.
However, since a mode deep inside the horizon oscillates with a frequency much larger than the Hubble scale, $f=c_sk/(2\pi a)\gg H$, 
a sharp transition from eMD to lRD enhances the amplitude of $\Phi$ substantially.
An exact solution of Eq.~\eqref{eq:curvatureeom} in lRD under the instantaneous evaporation approximation is given by
\begin{align}\label{eq:phi}
\Phi_{\rm lRD}(k\tau)=\frac{1}{c_sk\bar\tau}\left(C_1j_1(c_sk\bar \tau)+C_2y_1(c_sk\bar \tau)\right)\,;
\quad\bar\tau\equiv \tau-\tau_{\rm eva}/2,
\end{align}
where $c_s=1/\sqrt{3}$ is the sound velocity of radiation, 
$j_1$ and $y_1$ are respectively the spherical Bessel functions of the first and second kind of order $1$, and $C_1$ and $C_2$ 
are constants to be fixed by requiring continuity of the metric and its first time derivative at evaporation. 
The shift in the conformal time $\tau\to\bar\tau= \tau-\tau_{\rm eva}/2$ comes from the continuity of the background metric.
The constants $C_1$ and $C_2$ for modes which entered the horizon far before reheating, that is $k\gg k_{\rm eva}$ 
where $k_{\rm eva}={\cal H}_{\rm eva}$ is the scale that crosses the horizon at evaporation, read
\begin{equation}\label{eq:c1c2}
C_1=-\Phi_{\rm eMD}(k)\left(c_sk\tau_{\rm eva}/{2}\right)^{2}\cos(c_sk\tau_{\rm eva}/2)\,,
\quad
C_2=C_1 \tan(c_sk\tau_{\rm eva}/2)\,.
\end{equation}
See Appendix~\ref{App:Matchingconditions} for more details on the matching conditions.  {It is very important to note that although the PBH evaporation is almost instantaneous, the fact that it has a finite duration has a strong impact on scales which are much smaller than the evaporation rate \cite{Inomata:2020lmk}. We take into account this effect by introducing a suppression factor ${\cal S}_\Phi$, directly borrowing the results of \cite{Inomata:2020lmk}. We give the explicit expression for ${\cal S}_\Phi$ in the next section. We use the notation ${\cal S}_\Phi$ not to be confused with the isocurvature $S$.}

These results complete the whole evolution of the curvature perturbation from PBH formation through PBH evaporation 
until the standard radiation dominated universe. 
As clearly seen from Eqs.~\eqref{eq:phi} and \eqref{eq:c1c2}, the amplitude of $\Phi$ is enhanced by a factor of $O\left(k\tau_{\rm eva}\right)$.
This gives the time derivative ${\cal H}^{-1}\Phi'\sim (k\tau_{\rm eva})\Phi$ right after evaporation. 
Namely, the transition creates a large time derivative from zero,  
 ${\cal H}^{-1}\Phi'_{\rm eMD}(\tau_{\rm eva})=0$ to ${\cal H}^{-1}\Phi'_{lRD}(\tau_{\rm eva})\sim  (k\tau_{\rm eva})\Phi_{\rm eMD}$.
This induces a large amount of GWs, since the source is dominated by the term proportional to ${\cal H}^{-2}\Phi'^2$.
  Before going into the details of the induced GWs calculations, let us review the PBH reheating scenario and the scales involved.

\subsection{PBH reheating scenario\label{subsec:pbhreheating}}

\begin{figure}
\centering
\includegraphics[width=0.6\columnwidth]{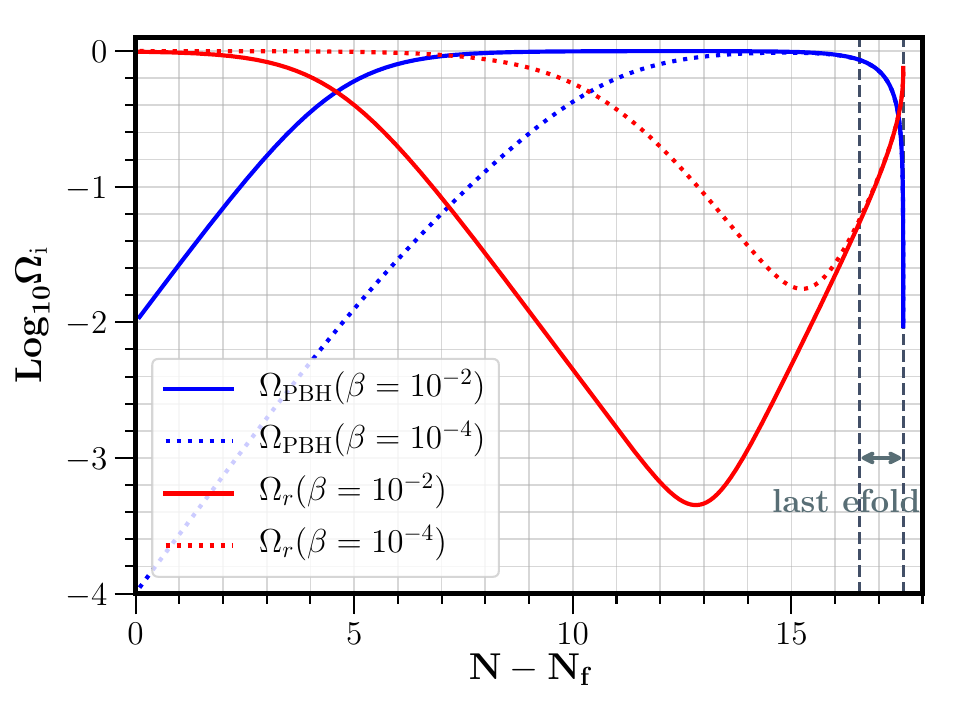}
\caption{Fractional energy density of the PBH gas (blue lines) and the radiation fluid (red lines) in terms of the number of e-folds 
from the time of PBH formation in the PBH reheating scenario. 
We chose $M=10\,{\rm g}$ and $\beta=10^{-2}$ (solid lines) and $\beta=10^{-4}$ (dotted lines). 
Although this particular range of $\beta$ is ruled out in Sec.~\ref{Sec:IIGWS}, it is a good illustrative example. 
Note how total evaporation is independent of $\beta$ and happens within less than $1/4$ of e-fold. 
Thus, evaporation is almost instantaneous. \label{fig:pbhreheating}}
\end{figure}

In this section, we mainly follow Ref.~\cite{Inomata:2020lmk} and we provide useful relations for the PBH reheating and 
isocurvature density fluctuations. We first proceed to study the relevant conditions and scales in the  PBH reheating scenario. 
The first condition is that the fraction of PBH at formation must be large enough to overcome the energy density of radiation 
before they completely evaporate. This can be estimated as follows. 
Evaporation by Hawking radiation decreases the PBH mass according to \cite{Kim:1999iv,Hooper:2019gtx}
\begin{align}\label{eq:hawkingevaporation}
\frac{dM_{\rm PBH}}{dt}=-\frac{A M_{\rm pl}^4}{M_{\rm PBH}^2}
\qquad {\rm where}\qquad
A=\frac{3.8\pi g_H(T_{\rm PBH})}{480}\,,
\end{align}
where $t$ is the cosmic time defined by $dt=ad\tau$, $g_H$ are the spin-weighted degrees of freedom and
\begin{align}
T_{\rm PBH}\equiv M_{\rm pl}^2/M_{{\rm PBH},f}\approx1.06\times 10^9{\rm GeV}\left(\frac{M_{\rm PBH,f}}{10^4{\rm g}}\right)^{-1}\,,
\end{align}
is the temperature associated to the PBH and we used that $M_{\rm pl}\equiv\sqrt{\hbar c/8\pi G}\approx 4.34\times 10^{-6}{\rm g}$. 
As we show later, we are only interested in black holes with $M_{\rm PBH,f}\ll 10^{11}{\rm g}$. 
Then we shall take that $g_H(T_{\rm PBH})\approx 108$ otherwise stated, assuming that the standard model of 
particle physics is valid up to $T_{\rm PBH}$ and no new physics are present. Integrating Eq.~\eqref{eq:hawkingevaporation} leads
 to a time dependent mass of the PBH as
\begin{align}\label{eq:MPBH(t)}
M_{\rm PBH}(t)\approx M_{{\rm PBH},f}\left(1-\frac{t}{t_{\rm eva}}\right)^{1/3}\,,
\end{align}
where we assumed that $t_{\rm eva}\gg t_f$ and we defined
\begin{align}\label{eq:teva}
t_{\rm eva}\approx\frac{M_{{\rm PBH},f}^3}{3AM_{\rm pl}^4}\,.
\end{align}
Within the approximation that $t_{\rm eva}\gg t_{\rm eq}>t_f$ we shall use that the Hubble parameter is well approximated by 
that of a pure MD, namely $H={\cal H}/a=2/(3t)$. 
In this case, the number of e-folds defined by $dN\equiv Hdt$ that the universe spends in the PBH dominated era is given by
\begin{align}\label{eq:efoldseva}
N_{\rm eva}-N_{\rm eq}\approx -\frac{2}{3}\ln\left[\frac{A}{2\pi\gamma\beta^2}\frac{M_{\rm pl}^2}{M_{{\rm PBH},f}^2}\right]
\approx 6.75 +\frac{4}{3}\ln\left[\frac{M_{{\rm PBH},f}}{10^4{\rm g}}\right]+\frac{4}{3}\ln\left[\frac{\beta}{10^{-7}}\right]\,.
\end{align}
To derive Eq.~\eqref{eq:efoldseva} we used the fact that $a_f/a_{\rm eq}=\beta$ and assumed that $\beta\ll 1$.
Requiring that $N_{\rm eva}>N_{\rm eq}$ in Eq.~\eqref{eq:efoldseva} sets a lower bound on $\beta$, which explicitly reads
\begin{align}\label{eq:betamin}
\beta>6.35\times10^{-10}\left(\frac{g_H(T_{\rm PBH})}{108}\right)^{1/2}
\left(\frac{\gamma}{0.2}\right)^{-1/2}\left(\frac{M_{{\rm PBH},f}}{10^4{\rm g}}\right)^{-1}\,.
\end{align}
We see that Eq.~\eqref{eq:betamin} is compatible with $\beta<1$ as long as $M_{{\rm PBH},f}>M_{\rm pl}$. 
We can also evaluate the reheating (evaporation) temperature of the universe in terms of the initial PBH mass as
\begin{align}
T_{\rm eva}\approx 2.76\times 10^4{\rm GeV}\left(\frac{M_{{\rm PBH},f}}{10^4{\rm g}}\right)^{-3/2}
\left(\frac{g_H(T_{\rm PBH})}{108}\right)^{1/2}\left(\frac{g_*(T_{\rm eva})}{106.75}\right)^{-1/4}\,,
\end{align}
where $g_*$ are the effective degrees of freedom (see Refs.~\cite{Husdal:2016haj,Saikawa:2018rcs} for a review 
on their dependence with the temperature). In order to have a successful BBN the reheating temperature of the universe must 
 be $T_{\rm eva}>4 {\rm MeV}$ \cite{Kawasaki:1999na,Kawasaki:2000en,Hannestad:2004px,Hasegawa:2019jsa}, 
 which in turn yields an upper bound on the PBH mass as
\begin{align}\label{eq:boundonmassBBN}
M_{{\rm PBH},f}<5\times 10^8{\rm g}\,.
\end{align}

At this point, we shall justify the sudden reheating assumption used throughout this work. To do this, we shall focus
 only on the evolution of the mean energy density of PBH, which decays according to (see Appendix~\ref{App:basicequations}) 
\begin{align}
\dot\rho_{\rm PBH}+\left(3H+\Gamma\right)\rho_{\rm PBH}=0\,,
\end{align}
where $\dot\,\equiv d/dt$ and we defined
\begin{align}\label{eq:Gamma}
\Gamma\equiv-\frac{d\ln M_{\rm PBH}}{dt}\,.
\end{align}
The equations of motion for the radiation fluid have the same factor $\Gamma\rho_{\rm PBH}$ but with opposite sign to 
ensure total energy conservation. Now, the sudden reheating approximation is valid if $\Gamma\gg H$. 
Since we know that initially the rate of evaporation $\Gamma/H$ is very small, the requirement of a sudden reheating is 
that $\Gamma/H$ reaches very large values in less than an e-fold. An explicit computation using Eqs.~\eqref{eq:MPBH(t)} 
and \eqref{eq:Gamma} yields
\begin{align}\label{eq:ratio}
\frac{\Gamma}{H}\approx \frac{1}{2}\left(\frac{t_{\rm eva}}{t}-1\right)^{-1}\,.
\end{align}
On one hand, we confirm that initially when $t\ll t_{\rm eva}$ we have that $\Gamma\ll H$, justifying that 
we neglect early energy transfer between components in Sec.~\ref{subsec:isoevolution}. 
On the other hand, a numerical inspection of \eqref{eq:ratio} tells us that $\Gamma/H\sim 0.1$ at $1$ e-fold before evaporation 
but quickly reaches $\Gamma/H>1$ at around $1/4$ of e-fold before evaporation (e.g. see Fig.~\ref{fig:pbhreheating}). 
Thus, PBHs evaporate almost instantaneously. 

{However, as we anticipated in the previous section, scales which have a time variation larger than the evaporation rate, that is modes with $k\gg \Gamma$ are very much affected by the finite duration \cite{Inomata:2020lmk}. The main reason is that, by the non-zero pressure, perturbations in the radiation fluid on scales $k\gg \Gamma$ are not effectively produced. Then the curvature perturbation is mainly sourced by the fluctuations in the PBH energy density. This means that, by the Poission equation, $k^2\Phi\sim a^2\rho_{\rm PBH}\delta_{\rm PBH}$ even when radiation dominates until the complete evaporation at $t=t_{\rm eva}$. Therefore, for $k\gg \Gamma$, $\Phi$ approximately decays as $\rho_{\rm PBH}\propto (1-t/t_{\rm eva})^{1/3}$ until $\Gamma\sim k$. This yields an additional relative suppression to $\Phi$ as
\begin{align}\label{eq:xi}
{\cal S}_\Phi(k)\equiv \frac{\Phi_{\rm lRD}}{\Phi^{\rm instant}_{\rm lRD}}\approx\left(\sqrt{\frac{2}{3}}\frac{k}{k_{\rm eva}}\right)^{-1/3} \,,
\end{align}
where the superscript ``instant'' refers to the instantaneous transition value. This was confirmed to be a good approximation by numerical calculations in Ref.~\cite{Inomata:2020lmk}. We expect this approximation to be even better for the very small scales we are considering.}

It is very important to note that the matching conditions at evaporation for the sudden reheating limit are only physically meaningful 
if done in the synchronous slicing comoving with the PBH fluid, in which the hypersurface of constant proper time spent in the PBH rest frame
 after formation coincides with the time slicing. Only in this gauge the evaporation takes places at the same time everywhere in the universe. 
Otherwise the evaporation process would depend non-locally in time.
 In Appendix~\ref{App:Matchingconditions} we show that continuity of the metric and its first time derivative at evaporation 
 in the synchronous comoving gauge leads to the continuity of the curvature perturbation in the Newtonian gauge $\Phi$ 
 and its first time derivative calculated in Section~\ref{subsec:isoevolution} and used in Section~\ref{Sec:IIGWS}.

\begin{figure}
\centering
\includegraphics[width=0.6\columnwidth]{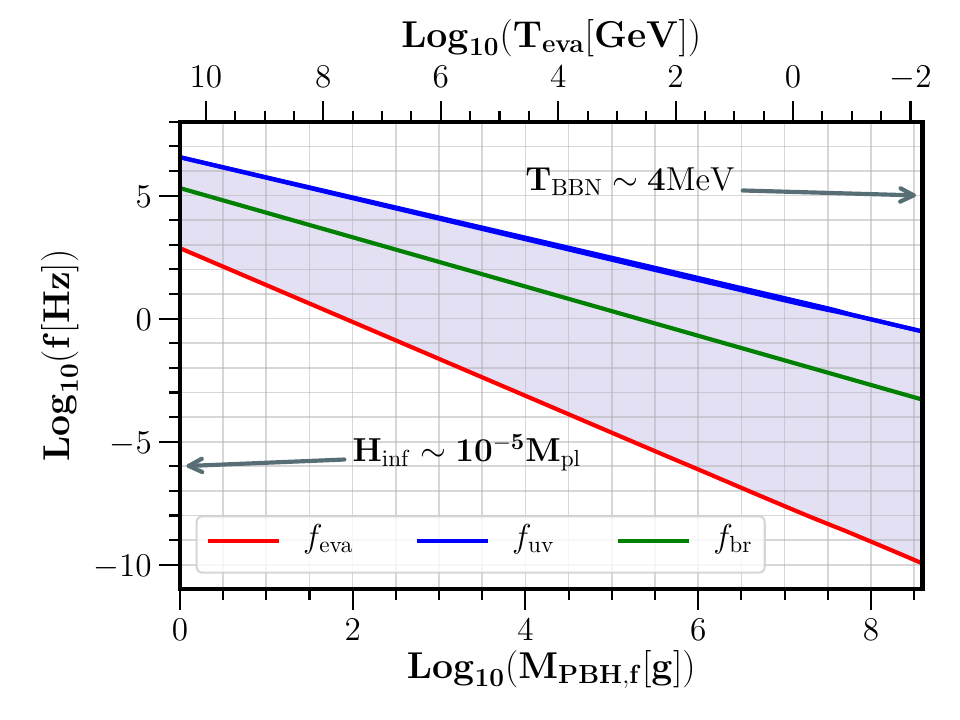}
\caption{Two characteristic frequencies of the PBH reheating scenario in terms of the PBH mass at formation (lower x axis) 
  and in terms of the reheating (evaporation) temperature (upper x axis). We show the frequency associated with
  the UV cut-off of the Poisson spectrum $f_{UV}$ in blue. The frequency corresponding to the comoving scale that 
  crosses the horizon at evaporation $f_{\rm eva}$ is shown in red. 
  Note that the ratio $f_{UV}/f_{\rm eva}$ depends only on $M_{{\rm PBH},f}$ and not on the fraction of PBHs at formation $\beta$. 
  The blue shaded region illustrates the extent in frequency of the GW spectrum derived in Sec.~\ref{Sec:IIGWS}. We also show the breaking frequency $f_{\rm br}$ where the GW spectrum changes slope given by Eq.~\eqref{eq:fbr}.
  See how $f_{UV}$, which also corresponds to the peak in the GW spectrum, falls within the observational range for the 
  largest masses, $M_{{\rm PBH},f}>10^{4}\,{\rm g}$. 
  In particular, it may enter the frequency range of LIGO frequency band \cite{LIGOScientific:2019vic}, LISA \cite{Audley:2017drz}, 
  Taiji \cite{Guo:2018npi}, Tianqin \cite{Luo:2015ght}, DECIGO \cite{Seto:2001qf,Yagi:2011wg}, AION/MAGIS \cite{Badurina:2019hst},
  and ET \cite{Maggiore:2019uih}. 
  We also illustrate the most left limit on the PBH mass corresponds to PBH formed right after a high scale inflation
   with $H_{\rm inf}\sim 10^{-5}M_{\rm pl}$. The most right limit is the requirement that for a successful BBN 
   one needs $T_{\rm eva}>4 {\rm MeV}$ \cite{Kawasaki:1999na,Kawasaki:2000en,Hannestad:2004px,Hasegawa:2019jsa}.\label{fig:f}}
\end{figure}

As we specified the details of the PBH reheating scenario and the matching conditions, it is useful to compare the relevant scales of the problem here.
They are the comoving wavenumbers that cross the horizon scale at the time of formation $k_f$, 
PBH-radiation equality $k_{\rm eq}$, and evaporation $k_{\rm eva}$, and the UV cut-off scale $k_{UV}$ beyond which 
the fluid description of PBHs ceases to be valid. We find the relations,
\begin{align}\label{eq:relatiosk}
\frac{k_{\rm eq}}{k_f}={\sqrt{2}\beta}\quad,\quad \frac{k_{\rm UV}}{k_f}=\left(\frac{\beta}{\gamma}\right)^{1/3}\,,
\quad \frac{k_{\rm eva}}{k_f}=\left(\frac{\beta A}{2\pi\gamma}\right)^{1/3}\left(\frac{M_{{\rm PBH},f}}{M_{\rm pl}}\right)^{-2/3}\,.
\end{align}
Note that the PBH reheating scenario is fully specified once $M_{{\rm PBH},f}$ and $\beta$ are given. 
From Eq.~\eqref{eq:relatiosk} we conclude that the hierarchy of the relevant scales reads
\begin{align}
k_f>k_{UV}>k_{\rm eq}>k_{\rm eva}\,.
\end{align}
It is also interesting to note that the ratio,
\begin{align}\label{eq:relatiosk2}
\frac{k_{UV}}{k_{\rm eva}}\approx {2.3}\times 10^6
 \left(\frac{g_H(T_{\rm PBH})}{108}\right)^{-1/3}\left(\frac{M_{{\rm PBH},f}}{10^4{\rm g}}\right)^{2/3}\,,
\end{align}
depends on the PBH mass at formation and does not involve the PBH fraction at formation. 
We show this in Fig.~\ref{fig:f}. 

 Here, let us explain the reason for the $\beta$-independence of the ratio $k_{UV}/k_{\rm eva}$.
  First, we note that the comoving wavenumber of the UV cut-off scale is determined by the mean separation distance between PBHs
 as $k_{UV}=d_f^{-1}a_f=n_f^{1/3}a_f$. This means $k_{UV}\propto \beta^{1/3}$. 
 As for the comoving wavenumber $k_{\rm eva}$,
 we note that the evaporation time of a black hole only depends on its mass and, hence, the Hubble parameter $H_{\rm eva}$
 and the PBH number density $n_{\rm eva}$ at evaporation are independent of the initial fraction of PBHs. 
 Then since $k_{\rm eva}=H_{\rm eva}(a_{\rm eva}/a_f)a_f$, and $a_{\rm eva}/a_f=(n_f/n_{\rm eva})^{1/3}$, it follows that
 $k_{\rm eva}\propto \beta^{1/3}$.  Thus, we find that both $k_{UV}$ and $k_{\rm eva}$ are proportional to $\beta^{1/3}$,
 implying the $\beta$-independence of the ratio $k_{UV}/k_{\rm eva}$.
 
 Furthermore, we note that the ratio \eqref{eq:relatiosk2} is always large, attaining $k_{UV}/k_{\rm eva}\sim 5000$ for the smallest masses 
 of $M_{{\rm PBH},f}\sim 1 {\rm g}$ which corresponds to $H_{\rm inf}\sim H_{f}\sim 10^{-5}M_{\rm pl}$. 
 We may obtain the value of any of the relevant scales using Eq.~\eqref{eq:relatiosk} and noting that
\begin{align}\label{eq:keva}
k_{\rm eva}\approx 4.7\times 10^{11}{\rm Mpc}^{-1}\left(\frac{g_H(T_{\rm PBH})}{108}\right)^{1/2}\left(\frac{g_*(T_{\rm eva})}{106.75}\right)^{1/4}\left(\frac{g_{*,s}(T_{\rm eva})}{106.75}\right)^{-1/3}\left(\frac{M_{{\rm PBH},f}}{10^4{\rm g}}\right)^{-3/2}\,,
\end{align}
where we also introduced $g_{*,s}$ as the entropic effective degrees of freedom. For instance, the frequency 
corresponding to the peak of the early isocurvature perturbation at $k_{UV}$ is given by 
\begin{align}
f_{UV}\approx 1.7\times 
10^{3}{\rm Hz}\left(\frac{g_H(T_{\rm PBH})}{108}\right)^{1/6}\left(\frac{g_*(T_{\rm eva})}{106.75}\right)^{1/4}
\left(\frac{g_{*,s}(T_{\rm eva})}{106.75}\right)^{-1/3}\left(\frac{M_{{\rm PBH},f}}{10^4{\rm g}}\right)^{-5/6}\,,
\end{align}
which interestingly may enter the current observational window of the LIGO/VIRGO collaboration. 
For $M_{{\rm PBH},f}\sim 1.6\times 10^6{\rm g}$, which roughly corresponds to $T_{\rm eva}\sim 15\,{\rm GeV}$ the peak frequency 
is around $f\sim 25\,{\rm Hz}$ where there are already constraints on the magnitude of the stochastic gravitational wave 
background (SGWB) \cite{LIGOScientific:2019vic}, concretely $\Omega_{\rm GW}<6\times 10^{-8}$. 
For $M_{{\rm PBH},f}\sim 10^8{\rm g}$ the peak frequency is around $f\sim 0.25\,{\rm Hz}$ and should be accessible 
by, e.g., DECIGO \cite{Seto:2001qf,Yagi:2011wg}. Note that for $M_{{\rm PBH},f}>2\times10^{4}\,{\rm g}$ 
the peak in the GW spectrum at $f_{UV}$ (see Sec.~\ref{Sec:IIGWS}) not only enters 
the LIGO frequency band \cite{LIGOScientific:2019vic} but also LISA \cite{Audley:2017drz}, Taiji \cite{Guo:2018npi}, 
Tianqin \cite{Luo:2015ght}, DECIGO \cite{Seto:2001qf,Yagi:2011wg}, AION/MAGIS \cite{Badurina:2019hst}, 
and ET \cite{Maggiore:2019uih}. For an illustration see Figs.~\ref{fig:f} and \ref{fig:gwscurves}.

\section{Early isocurvature induced gravitational waves\label{Sec:IIGWS}}

In this section, we shall estimate the amount of induced gravitational waves generated in the PBH reheating scenario 
and whether they may lead to substantial constraints on the fraction of PBHs at formation. 
The spectral density of induced GWs today may be expressed in terms of the GW spectral density deep inside the radiation dominated
 stage after evaporation, when it scales proportional to the total energy density of radiation, as \cite{Inomata:2019ivs}
\begin{align}\label{eq:OMGWs}
\Omega_{{\rm GW},0}(k)h^2=0.39\left(\frac{g_{*}(T_c)}{106.75}\right)^{-1/3}\Omega_{r,0}h^2\Omega_{{\rm GW},\rm c}(k)\,,
\end{align}
where $\Omega_{r,0}h^2\sim 4.18\times 10^{-5}$ is the energy density fraction of radiation evaluated today \cite{Aghanim:2018eyx}. 
In Eq.~\eqref{eq:OMGWs}, $\Omega_{{\rm GW},\rm c}(k)$ is evaluated at a time $\tau_{\rm c}$ during the lRD where the tensor modes $h_k$ propagate 
as a wave, that is $h'_k\approx k h$, and their spectral density remains constant, hence the subscript c. 
For the modes of interest we can assume that they already are propagating as a wave right after evaporation and reheating of the universe.
 At the time when GWs propagate as a wave we have 
\begin{align}\label{eq:OMGWC}
\Omega_{\rm GW,\rm c}(k)=\frac{k^2}{12{\cal H}_{\rm c}^2}\overline{{\cal P}_h(k,\tau_{\rm c})}\,,
\end{align}
where an overline denotes the time average over oscillations and we have used $h'\approx kh$. 
Note that from now on we set $M_{\rm pl}=1$ for simplicity.
 It should be noted that in order to take the oscillation average, the tensor modes $h_k$ must be deep inside the horizon 
 where they behave as a wave. It also implies that we must keep track of the GW evolution from PBH formation until 
 the radiation dominated universe after evaporation \cite{Inomata:2019ivs,Inomata:2020lmk}.

The power spectrum of tensor modes takes a compact form in terms of the initial scalar power spectrum squared. 
If we focus on the early isocurvature mode,\footnote{For the primordially adiabatic mode, which in this case would correspond 
to a peak in the primordial curvature power spectrum, one should replace 
\begin{align}
P_S(k)\to \left(\frac{3+3w}{5+3w}\right)^2P_{\cal R}(k)\,,
\end{align}
and would have to follow the evolution of the adiabatic curvature perturbation $\Phi$ until the later radiation dominated epoch. 
Here $P_{\cal R}(k)$ denotes the primordial curvature power spectrum.} 
the power spectrum may be calculated as \cite{Kohri:2018awv,Domenech:2019quo}
\begin{align}
\overline{{\cal P}_h(k,\tau)}=&
8\int_0^\infty dv\int_{|1-v|}^{1+v}du\left[\frac{4v^2-(1+v^2-u^2)^2}{4uv}\right]^2{\cal P}_{S}(kv){\cal P}_{S}(ku)\overline{I^2}(x,u,v)\,;
\label{eq:powerspectrum1}
\\
&I(x,u,v)=\int_{x_i}^x d\tilde x \,G(x,\tilde x) f(\tilde x, u, v)\,,
  \label{eq:kernel1}
\end{align}
where ${\cal P}_S(k)$ is the initial isocurvature perturbation spectrum, as presented in Eq.~\eqref{eq:PS}, 
$x\equiv k\tau$, $I(x,u,v)$ is the induced GWs kernel, $G(x,\tilde x)$ is the retarded Green function for the tensor modes 
 and $f(x,u,v)$ is proportional to the source term of tensor modes at second order in perturbation theory, 
which for the case of a radiation-matter universe reads
\begin{align}
f(x,u,v)=&
T_\Phi(ux)T_\Phi(vx)+\frac{3\rho}{2\rho_r}c_s^2\left(\frac{T_\Phi'(ux)}{{\cal H}}+T_\Phi(ux)\right)\left(\frac{T_\Phi'(vx)}{{\cal H}}+T_\Phi(vx)\right)
\nonumber\\
&\qquad+\frac{3}{2}a^2c_s^2\rho_mT_{V_{\rm rel}}(ux)T_{V_{\rm rel}}(vx)\,,
\label{eq:sourceterm}
\end{align}
where $T_\Phi(k,\tau)$ and $T_{V_{\rm rel}}(k,\tau)$ are the transfer functions for $\Phi$ and $V_{\rm rel}\equiv V_{\rm PBH}-V_r$, 
where $V_{\rm PBH}$ and $V_r$ are respectively the PBH gas and radiation spatial fluid velocities, defined by
\begin{align}
\Phi(k,\tau)=T_\Phi(k,\tau)S(k)\qquad{\rm and}\qquad V_{\rm rel}(k,\tau)=T_{V_{\rm rel}}(k,\tau)S(k)\,.
\end{align}
For more details see Appendix~\ref{App:basicequations}. 

Here let us comment on the contribution from the relative velocity term.
As a first approximation we may neglect the contribution from the relative velocity.
The reason is two-fold:
first, it is because the coefficient in front of their contribution to the GWs source is initially proportional to $\rho_m$ and later to $\rho_r$. 
Thus, deep inside both eRD and eMD, it becomes vanishingly small.
Second, it is because of the time evolution of the relative velocity \cite{Kodama:1986fg,Kodama:1986ud}.
  Initially it is negligible and eventually settles to a constant value proportional to $S/k_{\rm eq}$,
  as may be apparent from Eq.~\eqref{eq:eomsvrel} in the limits of superhorizon scales in eRD and for all scales in eMD. 
  However, by inspecting Eq.~\eqref{eq:sourceterm}, one might think
   that the contribution from the relative velocity could have some impact for scales $k\sim k_{\rm eq}$. 
  In the end though, we are mostly interested in scales close to the peak of the early isocurvature power spectrum at $k\sim k_{UV}\gg k_{\rm eq}$.
Furthermore, the GWs generated right after evaporation by the sudden jump in the time derivative of the curvature perturbation 
are the dominant contribution in the case of the eMD \cite{Inomata:2019ivs}. 
Thus, the GWs generated near the time of PBH-radiation equality are subdominant.

After a closer inspection of the Kernel, we may split Eq.~\eqref{eq:kernel1} into three contributions, schematically:
\begin{align}\label{eq:kernel2}
I(x,u,v)=I_{\rm eRD}(x,u,v,x_{\rm f},x_{\rm eq})+I_{\rm eMD}(x,u,v,x_{\rm eq},x_{\rm eva})+I_{\rm lRD}(x,u,v,x_{\rm eva})\,,
\end{align}
where we emphasized that there would be a dependence in $x_{\rm eq}$ and $x_{\rm eva}$ by requiring continuity of 
the tensor modes and their first derivative at the transitions. Out of these three contributions, the third one gives the largest
 contribution due to the enhanced amplitude of the oscillations of $\Phi$ \cite{Inomata:2019ivs} 
 (see the discussion below Eq.~\eqref{eq:c1c2}). Thus, for GWs on scales that entered the horizon prior to the beginning of
  the eMD we can safely assume that the dominant contribution comes from the time derivative of 
  the curvature perturbation at the lRD stage. Furthermore, for the case of a sudden transition one expects
  no correlation between modes generated in different eras \cite{Inomata:2019ivs}. 
  Therefore we may safely neglect the oscillation average of the product of different era contributions to the kernel, 
  e.g. $\overline{I_{\rm eMD}I_{\rm lRD}}\approx 0$. Bearing the above in mind, we can split the total tensor modes power spectrum as
\begin{align}\label{eq:phsplit}
\overline{{\cal P}_h(k,\tau)}\approx\overline{{\cal P}_{h,\rm eRD}(k,\tau)}+\overline{{\cal P}_{h,\rm eMD}(k,\tau)}+\overline{{\cal P}_{h,\rm lRD}(k,\tau)}\,.
\end{align}
The Kernel of the last piece in Eq.~\eqref{eq:phsplit} has the largest contribution from the terms in Eq.~\eqref{eq:sourceterm} 
which contain the largest number of time derivatives. From Eq.~\eqref{eq:c1c2} we see that, before any integration, 
the Kernel in the late RD has a factor $x_{\rm eva}^8$, where $x_{\rm eva}\equiv k\tau_{\rm eva}=2k/k_{\rm eva}$. 
This is a huge enhancement for the smallest scales since in the PBH reheating scenario $k_{UV}/k_{\rm eva}\gtrsim 5000$. 
We derive the analytical expressions for the kernel in Appendix~\ref{App:kernel}. The GW spectrum for the second contribution in Eq.~\eqref{eq:phsplit} is derived in Ref.~\cite{Papanikolaou:2020qtd}.

Now, we can estimate the spectral density of GWs of the resonance as follows. We first note that although the early
 isocurvature power spectrum peaks at $k\sim k_{\rm eq}$, the kernel \eqref{eq:irdapp} has an additional $k^8$ power 
 which shifts the peak of the integrand to the smallest scales at $k\sim k_{\rm UV}$. 
 Thus, we shall focus only on scales that entered the horizon before PBH-radiation equality, that is $k\gg k_{\rm eq}$. 
 In this case we have from Eq.~\eqref{eq:conversionfinal} that the transfer function during the transition is given by{
\begin{align}\label{eq:tmd}
T_{\Phi, lRD}(k)=\frac{3}{4}{\cal S}_\Phi^2(k)\left(\frac{k_{\rm eq}}{k}\right)^2\Theta(k-k_{\rm eq})\Theta(k_{UV}-k)\,,
\end{align} 
where we used the suppression due to the eRD from Eq.~\eqref{eq:conversionfinal} and the suppression factor $\xi$ due to the transition from Eq.~\eqref{eq:xi}.}
Note that by momentum conservation the UV cut-off leads to a cut-off of the GW spectrum at $k\leq2k_{UV}$
 and the IR cut-off to $k\geq 2k_{\rm eq}$. We derive the GW spectrum for $k<2k_{\rm eq}$ in Appendix~\ref{App:kernel}. 
 After plugging Eq.~\eqref{eq:tmd} and the dominant contribution to the kernel \eqref{eq:irdapp} in the last part of
  Eq.~\eqref{eq:phsplit} we find that {
\begin{align}\label{eq:phrld}
&\overline{{\cal P}_{h,\rm lRD}}(k,\tau,x\gg1)\approx
\frac{9c_s^4}{2^{18}\pi^2 x^2}\left(\frac{3}{2}\right)^{2/3}\left(\frac{k_{\rm eva}}{k_{UV}}\right)^{4/3}\left(\frac{k_{\rm eq}}{k_{UV}}\right)^8\left(\frac{k}{k_{\rm UV}}\right)^{-10/3} x_{\rm eva}^8
\nonumber\\
&\times\int_{k_{\rm eq}/k}^{k_{UV}/k} dv\int_{{\rm max}{(|1-v|,{k_{\rm eq}/k}})}^{{\rm min}({1+v,{k_{UV}/k}})}du
\left[\frac{4v^2-(1+v^2-u^2)^2}{4uv}\right]^2(uv)^{1/3}\,{\rm Ci}^2(|1-(u+v)c_s|x_{\rm eva}/2)\,.
\end{align}}
It should be noted that Eq.~\eqref{eq:phrld} is valid for any isocurvature model which has an early matter dominated stage
followed by an instantaneous reheating. {It is important to note that Ref.~\cite{Inomata:2019ivs} showed that even though evaporation has a finite duration, the instantaneous approximation together with the suppression factor $\xi$ matched well with the numerical results of the GW spectrum.}

\begin{figure}
\centering
\includegraphics[width=0.6\columnwidth]{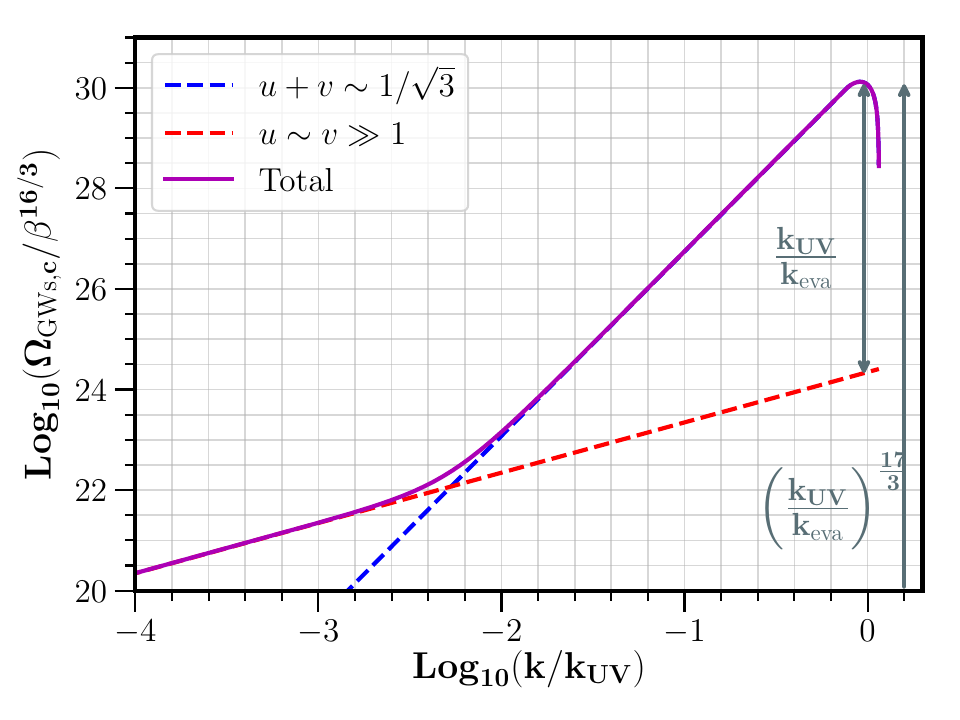}
\caption{GW spectrum induced by early isocurvature PBH fluctuations in terms of wavenumber for $M_{{\rm PBH},f}=10^4\,{\rm g}$. 
The spectrum has been normalized by a factor $\beta^{-16/3}$ so as to be independent of the fraction of PBHs at formation. 
The dashed blue line is the resonant contribution to the GW spectrum given in Eq.~\eqref{eq:GWsres}. 
The dashed red line is the large momentum contribution derived in Eq.~\eqref{eq:omegaLV}. 
The purple solid line is the total GW spectrum. The peak of the GW spectrum is located roughly at $k\sim k_{UV}$, 
has an amplitude proportional to {$(k_{UV}/k_{\rm eva})^{17/3}$} and is at least $k_{UV}/k_{\rm eva}$ higher than 
the IR tail of the spectrum, where $k_{UV}$ corresponds to the cut-off scale of the Poisson spectrum \eqref{eq:PS}. 
$k_{\rm eva}$ corresponds to the scale that crosses the horizon at evaporation \eqref{eq:keva}. 
The ratio $k_{UV}/k_{\rm eva}$ is found to be greater than $5000$. \label{fig:GWspectrum}}
\end{figure}

In order to perform such integral it is useful to change variables to $y\equiv \left((u+v)c_s-1\right){x_{\rm eva}}/{2}$ 
and $s\equiv u-v$, which has a Jacobian equal to $1/(c_sx_{\rm eva})$. 
Then, as a good approximation we evaluate the integrand at $y=0$ except for the divergent cosine integral \cite{Inomata:2019ivs}. 
By doing so we arrive at an approximate formula for the large $k$ part of the GWs spectrum deep inside the lRD \eqref{eq:OMGWC} given by {
\begin{align}\label{eq:GWsres}
\Omega_{\rm GW,c,\rm res}(k\gg k_{\rm eq})
\approx\frac{3c_s^{7/3}(1-c_s^2)^2}{2^{14}\pi}\left(\frac{9}{2}\right)^{1/3}\left(\frac{k}{k_{\rm eva}}\right)^{11/3}&\left(\frac{k_{\rm UV}}{k_{\rm eva}}\right)^2
\left(\frac{k_{\rm eq}}{k_{UV}}\right)^8\nonumber\\&\times\int_{-s_0(k)}^{s_0(k)} \frac{(1-s^2)^2}{(1-c_s^2s^2)^{5/3}}ds\,,
\end{align}}
where \cite{Inomata:2019ivs}
\begin{align}\label{eq:s0}
s_0(k)=\left\{
\begin{aligned}
&1\qquad &{k_{UV}}/{k}\geq\tfrac{1+c_s^{-1}}{2}\\
&2\tfrac{k_{UV}}{k}-c_s^{-1}\qquad & \tfrac{1+c_s^{-1}}{2}\geq\tfrac{k_{UV}}{k}\geq\tfrac{c_s^{-1}}{2}\\
&0\qquad &\tfrac{c_s^{-1}}{2}\geq\tfrac{k_{UV}}{k}
\end{aligned}
\right.\,.
\end{align}
The subscript res in Eq.~\eqref{eq:GWsres} stands for resonant contribution.
As an approximation in Eq.~\eqref{eq:GWsres} we have integrated $y$ from $-\infty$ to $\infty$ which yields 
an additional factor $\pi$ (see Eq.~\eqref{eq:ciint}). This comes at the cost of an $O(1)$ uncertainty 
which is not important nonetheless due to the large enhancement. Also, Eq.~\eqref{eq:s0} comes from the fact that the original 
variable $u$ has a cut-off which is ${\rm min}(1+v,k_{UV}/k)$, that is either by momentum conservation or by the cut-off 
of the scalar spectrum, and that we are imposing $y=0$ which implies $u=c_s^{-1}-v$. We see that the GW spectrum peaks 
at around the resonant scale $k\sim2c_sk_{UV}$, above which the spectrum is quickly suppressed until the cut-off 
at $k=2k_{UV}$. In fact, due to the dependence of $s_0(k)$ in Eq.~\eqref{eq:s0} we numerically find the peak at 
around $k\sim k_{UV}$, where the integral is approximately $1/2$. We may now evaluate the spectrum at the peak and 
use Eqs.~\eqref{eq:relatiosk} and \eqref{eq:relatiosk2} to express it in terms of the PBH scenario parameters, namely{
\begin{align}\label{eq:gwspeak}
\Omega_{\rm GW,c,\rm res}(k\sim k_{UV})
&\approx \frac{1}{24576\pi\,2^{1/3}\sqrt{3}}\left(\frac{k_{UV}}{k_{\rm eva}}\right)^{17/3} \left(\frac{k_{\rm eq}}{k_{UV}}\right)^8
\nonumber\\&\approx 10^{30}\beta^{16/3}\left(\frac{\gamma}{0.2}\right)^{8/3}\left(\frac{g_H(T_{\rm PBH})}{108}\right)^{-17/9}
\left(\frac{M_{{\rm PBH},f}}{10^4{\rm g}}\right)^{34/9}\,.
\end{align}}
We show the total spectrum of GWs in Fig.~\ref{fig:GWspectrum}. See how the infrared tail Eq.~\eqref{eq:omegaLV} derived 
in Appendix~\ref{App:kernel} is suppressed by at least a factor $k_{UV}/k_{\rm eva}$. 
Such a suppression deems the infrared tail unreachable to future GWs detectors. See Fig.~\ref{fig:gwscurves}. {Nevertheless, we may hope to see the knee of the GW spectrum. Using the results of Appendix~\ref{App:kernel}, we find that the breaking frequency of the GW spectrum is located at
\begin{align}\label{eq:fbr}
f_{\rm br}\approx 10 \,{\rm Hz}\left(\frac{g_H(T_{\rm PBH})}{108}\right)^{7/24}\left(\frac{g_*(T_{\rm eva})}{106.75}\right)^{1/4} \left(\frac{g_{*,s}(T_{\rm eva})}{106.75}\right)^{-1/3} \left(\frac{M_{\rm PBH}}{10^4\,{\rm g}}\right)^{-13/12}\,,
\end{align}
which might enter the observational window (see Fig.~\ref{fig:f}).
}

\begin{figure}
\centering
\includegraphics[width=0.6\columnwidth]{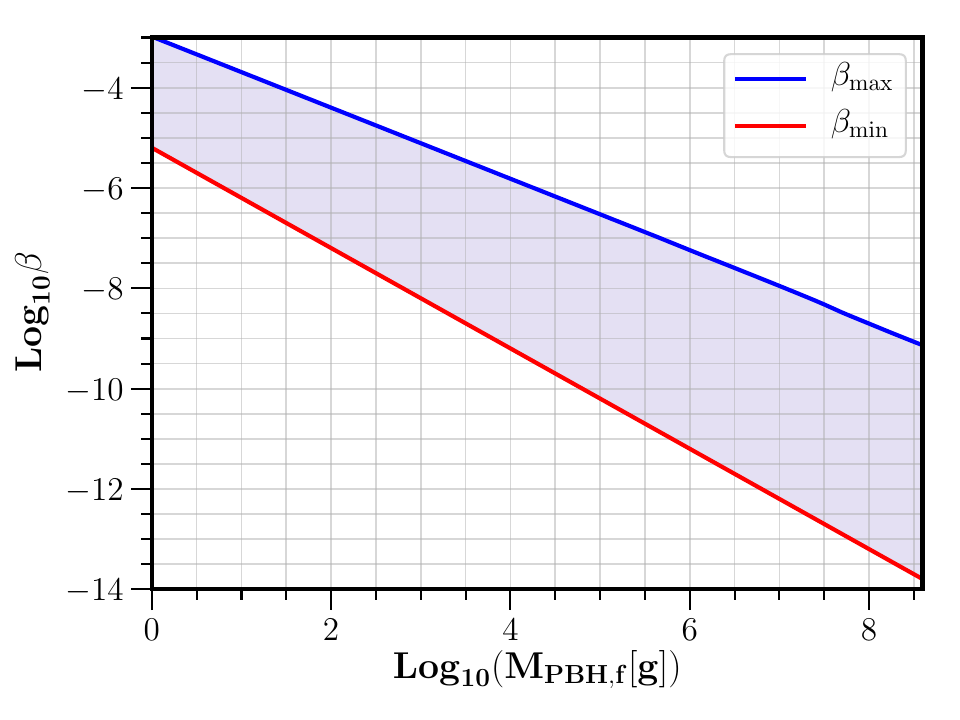}
\caption{Allowed parameter space for $\beta$ as a function of $M_{{\rm PBH},f}$. 
The lower bound $\beta_{\rm min}$ shown in red comes from requiring PBH domination prior to evaporation \eqref{eq:betamin}. 
The upper bound show in blue \eqref{eq:betamax} comes from the BBN bound on the effective number of species, which sets 
an upper bound to the GW energy density fraction. The blue shaded region shows the allowed parameter region for $\beta$ 
in the PB reheating scenario. See how it is only restricted to roughly 2 orders of magnitude. \label{fig:beta}}
\end{figure}

Any additional relativistic species during lRD at BBN should contribute no more than $20\%$ of the total 
energy density \cite{Aghanim:2018eyx}.\footnote{The constraint on the energy density ratio of gravitational waves at the time of BBN is \cite{Caprini:2018mtu}
\begin{align}
\Omega_{\rm GW,BBN}<\frac{7}{8}\left(\frac{4}{11}\right)^{4/3}\Delta N_{\rm eff}\sim 0.05\,,\nonumber
\end{align}
where we used $\Delta N_{\rm eff}\lesssim0.2$ from the latest 2018 Planck results \cite{Aghanim:2018eyx}.
} 
Thus, requiring that $\Omega_{\rm GW,BBN}\approx 0.39\,\Omega_{\rm GW,c,\rm res}<0.05$ leads
 to\footnote{Although the BBN bound is an integrated constraint, i.e. a constraint on $\int d\ln k\,\Omega_{\rm GW}(k)$, 
  our GW spectrum is very peaked and, therefore, considering only the peak of the spectral density is good enough for our purposes.}{
\begin{align}\label{eq:betamax}
\beta<1.1\times 10^{-6}\left(\frac{\gamma}{0.2}\right)^{-1/2}\left(\frac{g_H(T_{\rm PBH})}{108}\right)^{17/48}
\left(\frac{g_*(T_{\rm eva})}{106.75}\right)^{1/16}\left(\frac{M_{{\rm PBH},f}}{10^4{\rm g}}\right)^{-17/24}\,.
\end{align}
Combining this upper bound and the lower bound derived in Eq.~\eqref{eq:betamin}, 
we find that the range of the initial PBH fraction is restricted to
\begin{align}\label{eq:betarange}
1.1\times 10^{-6}\left(\frac{M_{{\rm PBH},f}}{10^4{\rm g}}\right)^{-17/24}
\gtrsim\beta\gtrsim
6.4\times10^{-10}\left(\frac{M_{{\rm PBH},f}}{10^4{\rm g}}\right)^{-1}\,.
\end{align}
This is a rather strong constraint on the initial fraction $\beta$ of PBHs and spans only for $2$ orders of magnitude depending
on the PBH mass. We show the detailed bounds as a function of PBH mass in Fig.~\ref{fig:beta}. 
 For example, for $M_{{\rm PBH},f}\sim 1\,{\rm g}$ we find $7.8\times10^{-4}\gtrsim\beta\gtrsim6.3\times 10^{-6}$ 
 while for $M_{{\rm PBH},f}\sim 5\times 10^8\,{\rm g}$ we find $5\times 10^{-10}\gtrsim\beta\gtrsim1.2\times 10^{-14}$. 
 Thus, we conclude that PBHs cannot dominate the universe at formation and the PBH fraction at formation must be smaller 
 than $10^{-4}$ for $M\sim 1{\rm g}$ and $10^{-10}$ for the largest masses allowed by the BBN constraint on the reheating 
 temperature \eqref{eq:boundonmassBBN}. }
 
 It should be noted from Eq.~\eqref{eq:gwspeak} that the upper bound on $\beta$ \eqref{eq:betamax}  depends on $\Omega_{GW}^{3/16}$. 
 This implies that in order to rule out the PBH reheating scenario by requiring  the upper bound \eqref{eq:betamax} to be 
 smaller than the lower bound \eqref{eq:betamin}  {
 requires  $\Omega^{\rm constraint}_{GW,BBN}<10^{-19}(M_{{\rm PBH},f}/10^4\,{\rm g})^{-14/9}$ or 
 translated to the GW spectral density today yields $\Omega^{\rm constraint}_{GW,0}h^2<10^{-24}(M_{{\rm PBH},f}/10^4\,{\rm g})^{-14/9}$. }
Unfortunately, this upper bound on the spectral density of GWs is out of reach of current and (foreseeable) future GW detectors and, 
therefore, PBH reheating cannot be completely ruled out by means of the induced GWs from the early isocurvature perturbation. 
  For example, using the recent LIGO bounds on the SGWB \cite{LIGOScientific:2019vic} we have for 
  $M_{{\rm PBH},f}\sim 10^6\,{\rm g}$ that {$\beta<3\times 10^{-8}$} which is only an improvement by a factor 2 
  with respect to Eq.~\eqref{eq:betamax}.
 The optimistic DECIGO sensitivity curves would yield {$\beta<10^{-11}$} for $M_{{\rm PBH},f}\sim 5\times 10^8 \,{\rm g}$, {
 $10$ times better than $\eqref{eq:betamax}$ but still a factor $10^3$ from the lower bound \eqref{eq:betamin}. }
 Most interestingly though, PBH reheating could be found by ET and DECIGO respectively for PBH masses in 
 the range $10^8\,{\rm g}>M_{{\rm PBH},f}>2\times 10^4\,{\rm g}$ and $5\times 10^8\,{\rm g}>M_{{\rm PBH},f}>5\times 10^6\,{\rm g}$, e.g. see Fig.~\ref{fig:gwscurves}.

We may include the effects of clustering by including a higher order $k$ dependence on scales $k>k_{\rm cluster}$. 
In that case we have to shift our estimate to{
\begin{align}\label{eq:GWsrescluster}
&\Omega_{\rm GW,c,\rm res}^{{\rm cluster}}(k\gg k_{\rm cluster})\approx F_{\rm cluster}^2 4^{-n}c_s^{7/3-2n} \left(\frac{9}{2}\right)^{1/3}
\frac{3(1-c_s^2)^2}{2^{14}\pi\bar x^2}
\nonumber\\
&\times\left(\frac{k}{k_{\rm eva}}\right)^{11/3}\left(\frac{k}{k_{\rm cluster}}\right)^{2n}\left(\frac{k_{\rm eq}}{k_{\rm eva}}\right)^2
\left(\frac{k_{\rm eq}}{k_{UV}}\right)^6\int_{-s_0(k)}^{s_0(k)} {(1-s^2)^2}{\left(1-c_s^2s^2\right)^{n-5/3}}ds\,.
\end{align}}
Since the integral over $s$ results in hypergeometric functions we have a rough estimate as follows. 
We first consider only the case $s_0=1$ in Eq.~\eqref{eq:s0} since any deviation yields an $O(1)$ factor which is not important. 
Then we evaluate the spectrum at $k\sim k_{UV}$ where the peak is expected, which yields{
\begin{align}\label{eq:gwspeakcluster}
\Omega^{\rm cluster}_{\rm GW,c,\rm res}(k\sim k_{UV})\approx
10^{30}\beta^{16/3}\tilde F_{\rm cluster}^2\left(\frac{k_{UV}}{k_{\rm cluster}}\right)^{2n}
\left(\frac{\gamma}{0.2}\right)^{8/3}\left(\frac{g_H(T_{\rm PBH})}{108}\right)^{-17/9}\left(\frac{M_{{\rm PBH},f}}{10^4{\rm g}}\right)^{34/9}\,.
\end{align}}
where the precise value of $\tilde F_{\rm cluster}=\tilde F_{\rm cluster}(F_{\rm cluster},n,c_s)$ is given in Appendix~\ref{App:kernel}. 
The presence of clustering might set tighter constraints on $\beta$ given by{
\begin{align}\label{eq:betamaxcluster}
\beta<1.1\times 10^{-6}\tilde F_{\rm cluster}^{-3/8}\left(\frac{k_{\rm cluster}}{k_{UV}}\right)^{3n/8}\left(\frac{\gamma}{0.2}\right)^{-1/2}
\left(\frac{g_H(T_{\rm PBH})}{108}\right)^{17/48}\left(\frac{M_{{\rm PBH},f}}{10^4{\rm g}}\right)^{-17/24}\,.
\end{align}}
For example, if we consider $n=1$ and we take a large clustering $\tilde F_{\rm cluster}^2\approx 4/5 F_{\rm cluster}^2\sim 1$, 
the parameter space of PBH reheating is non-vanishing if{
\begin{align}
\frac{k_{\rm cluster}}{k_{UV}}\gtrsim2\times 10^{-9}\left(\frac{M_{{\rm PBH},f}}{10^4{\rm g}}\right)^{-7/9}\,,
\end{align}}
where we imposed that the upper bound \eqref{eq:betamaxcluster} is larger than the minimum value of $\beta$ \eqref{eq:betamin}. We see that the PBH isocurvature induced GWs is an interesting way to constrain the amount of clustering in the PBH reheating scenario. 

\begin{figure}
\centering
\includegraphics[width=0.6\columnwidth]{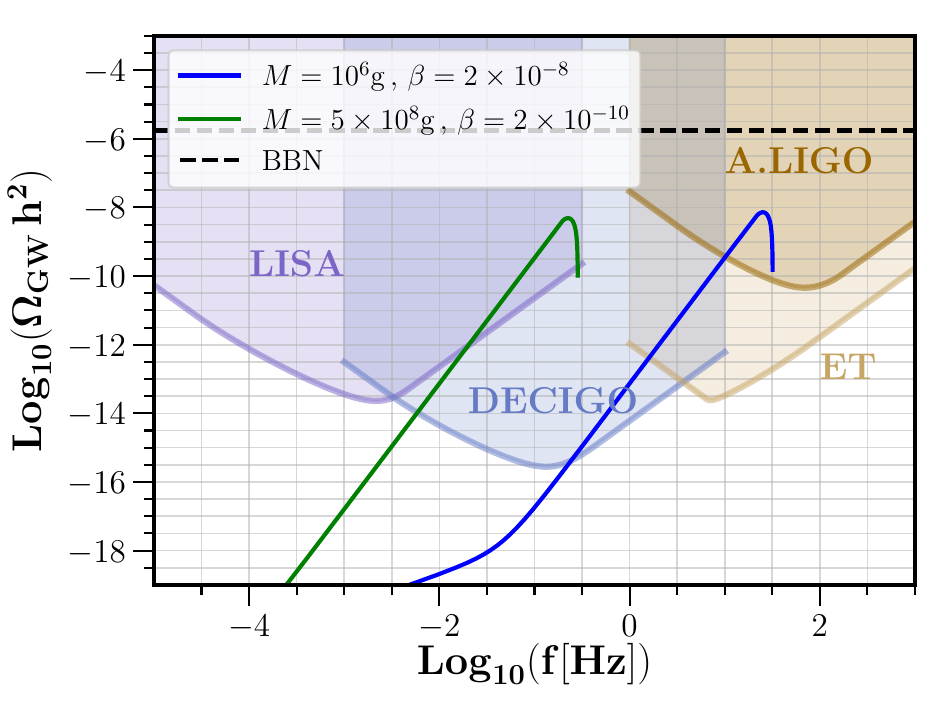}
\caption{ GW spectral density induced by early isocurvature fluctuations in the PBH reheating scenario in terms of frequency. 
The blue and green solid lines respectively correspond to {($M_{{\rm PBH},f}= 10^6\,{\rm g}$ , $\beta=2\times10^{-8}$) 
and ($M_{{\rm PBH},f}=5\times 10^8\,{\rm g}$ , $\beta=2\times10^{-10}$)}. 
Note that $M_{{\rm PBH},f}$ and $\beta$ are the only two free parameters in the model. 
We also show the power-law integrated sensitivity curves for LISA, DECIGO, ET and Advanced LIGO \cite{Thrane:2013oya}.
 \label{fig:gwscurves}}
\end{figure}

\section{Discussion and conclusions\label{sec:conclusions}}

Primordial black holes are very interesting objects which could provide information about any physics prior to the big bang nucleosynthesis, 
including inflation \cite{Sasaki:2018dmp}. Accompanying the formation of PBHs is the generation of induced gravitational waves which, 
due to the large size of the fluctuations that collapse into PBHs, have an amplitude large enough to enter the current and future 
GW observational window 
\cite{tomita,Matarrese:1992rp,Matarrese:1993zf,Matarrese:1997ay,Noh:2004bc,Carbone:2004iv,Nakamura:2004rm,Ananda:2006af,Baumann:2007zm,Osano:2006ew}.
These induced GWs are mainly sourced by any curvature perturbation. 

There are three distinguished stages in the PBH reheating scenario. Initially the universe is radiation dominated, which we called
eRD. Then PBHs are formed and they begin to dominate the universe. The PBH dominated stage is called eMD.
After PBHs have dominated the universe, they evaporate due to Hawking radiation. The evaporation process occurs almost instantaneously,
and the universe becomes radiation dominated again. This last stage is called lRD. During these three stages, we have two relevant sources of induced GWs. One is a primordial curvature perturbation which 
is responsible for the formation of PBHs. The other is the density fluctuations that arise from the formation of PBHs. 
Since PBH formation is a rare event, their spatial distribution follows the Poisson statistics in the standard case. 
Hence they will have a Poisson spectrum with UV cutoff at the scale of the mean distance between PBHs \cite{Papanikolaou:2020qtd}.
Assuming PBHs are formed during eRD,
PBHs constitute only a tiny fraction of the energy density at the time of formation. 
Hence the PBH density fluctuations are isocurvature in nature, which we call the early isocurvature perturbation.
It is converted into a curvature perturbation after PBHs has dominated the universe. 
Thus, for a peaked primordial curvature power spectrum, which is responsible for the PBH formation,
GWs are sourced first by the primordial curvature perturbation during the eRD era and later by the early isocurvature perturbation
during the eMD and lRD. Since PBH evaporation occurs in a time scale shorter than the Hubble time,
an instantaneous reheating approximation is valid, and the induced GWs are very much enhanced
due to the large amplitude of the oscillations of the curvature perturbation after the sudden evaporation \cite{Inomata:2019ivs,Inomata:2020lmk}.
Thus the most important contribution to the early isocurvature induced GWs comes from the stage right after PBH reheating.

In this paper, we extended the work of Ref.~\cite{Papanikolaou:2020qtd} in several ways. We calculated the dominant contribution to the induced GWs due to the early isocurvature perturbation in the PBH reheating scenario for 
a monochromatic mass function, or alternatively a very sharply peaked primordial curvature power spectrum. We derived the shape of the peak in the GW spectrum due to the sudden transition from matter to radiation 
domination in Eqs.~\eqref{eq:GWsres} and the infrared tail in Eqs.~\eqref{eq:omegaLV} and \eqref{eq:omegaLV2}. 
We also noticed that the ratio of the scale that crosses the horizon at evaporation $k_{\rm eva}$ 
and the UV cut-off scale of the Poisson spectrum $k_{UV}$ is independent of the fraction of PBHs at formation $\beta$. 
We found that this ratio is always fairly large, at least $k_{UV}/k_{\rm eva}\gtrsim 5000$. 
 This implies that the peak of the induced GW spectrum \eqref{eq:gwspeak} which is proportional to {$\beta^{16/3}(k_{UV}/k_{\rm eva})^{17/3}$}
 can have a huge amplitude unless $\beta$ is extremely small. 
 The amplitude of the IR tail \eqref{eq:omegaLV} is negligible and is at least  suppressed a factor $(k_{\rm eva}/k_{UV})$ with respect to the peak.
 Using the BBN constraints on the effective number of degrees of freedom, we found that the fraction of PBH at formation $\beta$ is
  constrained to be a tiny fraction of the total energy density. 

Together with the requirement that the PBH fraction is large enough so that PBH eventaully dominate before evaporation, 
we find in Eq.~\eqref{eq:betarange} that the allowed range of $\beta$ only spans for {two to four} orders of magnitude. 
For example, for $M_{{\rm PBH},f}\sim 1{\rm g}$ (if PBHs form right after a high scale inflation) 
we find {$8\times10^{-4}\gtrsim\beta\gtrsim6\times 10^{-6}$ while for $M_{{\rm PBH},f}\sim 5\times 10^8{\rm g}$ 
(if evaporation occurs right at the BBN bound of $T_{\rm eva}\sim 4\,{\rm MeV}$) 
we find $5\times 10^{-10}\gtrsim\beta\gtrsim 10^{-14}$.}
It should be noted that the upper bound \eqref{eq:betamax} on $\beta$ depends on the constraint of the amplitude of the SGWB to 
the power of $-3/16$, i.e $\propto\Omega_{\rm GW}^{-3/16}$. Thus, it will be difficult to improve the upper bound on $\beta$ by 
future constraints on the SGWB. In fact, PBH reheating is unlikely to be ruled out by the early isocurvature induced GWs as 
one would need a sensitivity on $f\sim {\rm Hz}$ around {$\Omega_{\rm GW}\sim 10^{-24}$.} On the other hand, for PBHs with 
initial masses $M_{{\rm PBH},f}\lesssim 2\times 10^4$ the peak of the induced GW spectrum could 
enter the LIGO \cite{LIGOScientific:2019vic}, LISA \cite{Audley:2017drz}, Taiji \cite{Guo:2018npi}, Tianqin \cite{Luo:2015ght}, 
DECIGO \cite{Seto:2001qf,Yagi:2011wg}, AION/MAGIS \cite{Badurina:2019hst} and ET \cite{Maggiore:2019uih} frequency range. 
It is exciting that future GW detectors might be able to probe the PBH reheating scenario.

We also considered in a phenomenological way that PBHs might cluster below a given scale, which we called $k_{\rm cluster}<k_{UV}$, 
due to a scale dependent non-gaussianity in the primoridal curvature perturbation. The clustering will cause the spectrum of
 PBH density fluctuations to grow on the smallest scales and, therefore, the enhancement of the induced GWs will be more
  pronounced than  \eqref{eq:gwspeakcluster}. 
 Thus, the magnitude of clustering and/or the scale of clustering are respectively constrained from above and below. 
A concrete model of PBH formation with cluster can thus be constrained by Eq.~\eqref{eq:betamaxcluster}. We mention that, although we focused on the particular, PBH reheating scenario,
our methodology presented in this paper can be applied to any early isocurvature model with an eMD.

In this paper we did not consider the evolution of  an early adiabatic perturbation, i.e., the primordial curvature perturbation,
and GWs induced by it. These were well studied in, e.g., Refs.~\cite{Kohri:2018awv,Inomata:2019zqy,Inomata:2019ivs,Inomata:2020lmk}. 
Here let us explain why the induced GWs by the primordial curvature perturbation are not important in the PBH reheating scenario
as far as observational constraints are concerned.
First, for a sharp peak in the primordial curvature power spectrum with amplitude ${\cal A}_{\cal R}$, 
the peak of the induced GW is at $k\sim 2k_f/\sqrt{3}$ with 
amplitude $\Omega_{\rm GW,c}^{e{\rm ad}}\sim 21 {\cal A}_{\cal R}^2$ \cite{Kohri:2018awv,Cai:2018dig}, 
where $k_f$ is the scale corresponding to the horizon at formation and {\it e}ad refers to early adiabatic mode. We show in App.~\ref{App:basicequations} Eq.~\eqref{eq:curvaturefinal2} that in the case of a sharp peak the contribution to the GWs spectrum induced by the primordial curvature perturbation right after evaporation is negligible. 
Second, for the allowed range of $\beta$ \eqref{eq:betarange}, $k_f\gg k_{UV}\gg k_{\rm eq}$ and $k_f$ is at least $10^4$ times bigger than $k_{UV}$, 
which is far from any observational window. This also means that most of the GWs induced by the primordial curvature perturbation are 
generated in RD. We estimate the amplitude of this induced GW spectrum as follows. First, a PBH initial fraction in the range $10^{-4}>\beta>10^{-14}$ implies that $0.21>{\cal A_R}>0.04$ \cite{Sasaki:2018dmp}. 
Second, the GW spectral density is redshifted by a factor $a^{-1}$ during eMD and is therefore suppressed
by a factor $a_{\rm eva}/a_{\rm eq}$ at the time of evaporation, which is at least $10^{-2}$ (see Eq.~\eqref{eq:efoldseva}). 
Collecting all the above estimate we obtain $\Omega^{e{\rm ad}}_{\rm GW,c}\lesssim8\times 10^{-3}$, 
which is an order of magnitude below the BBN bound.

It is important to note that the constraints derived in this work should be taken as rough bounds on $\beta$ and that the calculation 
of the induced GWs could be improved in several ways. One is the lower bound on $\beta$ given by \eqref{eq:betamin}. 
It is derived by requiring that PBH-radiation equality occurs before complete evaporation. Hence if $\beta$ approaches the lower 
bound \eqref{eq:betamin}, the equality and evaporation will become too close and the conversion of the perturbation from 
isocurvature to curvature will be less efficient. This will suppress the generation of  induced GWs. 
Another is the upper bound on $\beta$, \eqref{eq:betamax}. We assumed an instantaneous evaporation, 
which is good enough for our purpose, but accounting for the finite duration of reheating might also slightly 
suppress the induced GW spectrum \cite{Inomata:2020lmk}. 
Although we do not expect this to change much the upper bound on $\beta$ as the amplitude of the GW spectrum depends 
on a large power of $\beta$, it is desirable to derive the bound without the instantaneous reheating approximation. 

Another, and perhaps the most important issue is the fact that the density contrast of the PBH fluctuations, defined as $\delta\rho_{\rm PBH}/\rho_{\rm PBH}$, exceeds unity at evaporation. Concretely, using the bounds on $\beta$ Eq.~\eqref{eq:betarange} we find that $0.5\lesssim\delta\rho_{\rm PBH}/\rho_{\rm PBH}\lesssim 90 (M_{\rm PBH}/10^{4}{\rm g})^{1/6}$. This fact was already noted in Ref.~\cite{Papanikolaou:2020qtd}. Thus, we may regard our bounds on $\beta$ as conservative estimates.
Finally, we should relax the assumption on the PBH mass spectrum that it is monochromatic.
It would certainly affect the spectrum of early isocurvature perturbations as well as the reheating time-scale.
These are important issues but they are out of the scope of the current paper. We leave the cases of non-linear contributions from the large density contrast and an extended mass spectrum for future work.

\section*{Acknowledgments} 
We would like to thank K.~Inomata, D.~Langlois, T.~Papanikolaou and V.~Vennin for useful correspondence. G.D. would also like to thank J.~Rubio for useful discussions on isocurvature perturbations. G.D. as a Fellini fellow was supported by the European Union’s Horizon 2020 research and innovation programme under the Marie Sk{\l}odowska-Curie grant 
agreement No 754496. 
M.S. was supported in part by the JSPS KAKENHI Nos.~19H01895, 20H04727 and 20H05853. C.L. is supported by the grant No. UMO-2018/30/Q/ST9/00795 from the National Science Centre (Poland). 
Calculations of cosmological perturbation theory at second order were checked using the \texttt{xPand (xAct) Mathematica} package.

\appendix

\section{Basic equations\label{App:basicequations}}
For completeness we show in this appendix the basic equations for cosmological perturbation in a matter-radiation dominated universe. 
We perturb the metric as follows:
\begin{align}
ds^2=a^2(\tau)\left[-(1+2\Psi)d\tau^2+(\delta_{ij}+2\Phi\delta_{ij}+h_{ij})dx^idx^j\right]\,,
\end{align}
where $\Psi$ and $\Phi$ are scalar degrees of freedom and $h_{ij}$ are the transverse-traceless part of the metric. 
Then, we have the matter and radiation fluids which its energy momentum tensor is respectively given by
\begin{align}
T_{m\mu\nu}&=\rho_m u_{m\mu}u_{m\nu}\,,\\
T_{r\mu\nu}&=(\rho_r+p_r)u_{r\mu}u_{r\nu}+p_r g_{\mu\nu}\,,
\end{align}
where the 4-velocities are normalised as $g^{\mu\nu}u_{m\mu}u_{m\nu}=g^{\mu\nu}u_{r\mu}u_{r\nu}=-1$. 
Since PBH decay into radiation and evaporate, there is an energy transfer between components given by
\begin{align}
\nabla_\mu T^{m\mu\nu}&=-Q^\nu\,,\\
\nabla_\mu T^{r\mu\nu}&=Q^\nu\,,
\end{align}
where total energy is conserved and we defined
\begin{align}\label{eq:Q}
Q^\nu=Qu_m^\nu\quad{,}\quad Q\equiv-n_{\rm PBH}u^\mu_m\partial_\mu M_{\rm PBH}
\end{align}
and $n_{\rm PBH}=\rho_{\rm PBH}/M_{\rm PBH}$. Since $Q^\nu$ is due to the PBH evaporation it is useful to write it in terms of the PBH 4-velocity.

\subsection{Background}
At the background level, we have $u_{m0}=u_{r0}=-a$ and $u_{mi}=u_{ri}=0$. The Friedman and energy conservation equations read
\begin{align}
3{\cal H}^2&=a^2(\rho_m+\rho_r)\equiv a^2\rho\,,\\
{\cal H}^2+2{\cal H}'&=-\frac{1}{3}a^2\rho_r\,,\\
\rho_m'+3{\cal H}\rho_m&=-aQ\,,\\
\rho_r'+4{\cal H}\rho_r&=aQ\,.
\end{align}
When $Q=0$ an exact solution to the scale factor is given by
\begin{align}
a(\tau)/a_{\rm eq}=2\xi+\xi^2
\end{align}
where $\xi\equiv\tau/\tau_*$ and $(\sqrt{2}-1)\tau_*=\tau_{\rm eq}$.

\subsection{First order Newtonian Gauge}
Up to the linear perturbation level, we have $u_{m0}=u_{r0}=-a\left(1+\Psi\right),~u_{mi}=a\partial_iV_m$ and $u_{ri}=a\partial_iV_r$, 
where we have neglected the transverse modes for $u_m$ and $u_r$. 
Here we derive the equations of motion for linear perturbations. Starting with Einstein Equations we find from the traceless $ij$ component:
\begin{align}
\Phi+\Psi=0\,.
\end{align}
For simplicity we use $\Psi=-\Phi$ in all forthcoming equations. From the $00$, $0i$ and $ij$ trace components we respectively obtain:
\begin{align}
&6{\cal H}\Phi'+6{\cal H}^2\Phi-2\Delta\Phi=a^2(\delta\rho_m+\delta\rho_r)\equiv a^2\delta\rho\,,\\
&\Phi'+{\cal H}\Phi=\frac{1}{2}a^2\left(\rho_mV_m+\frac{4}{3}\rho_rV_r\right)\equiv\frac{1}{2}a^2\rho V\,,\\
&\Phi''+3{\cal H}\Phi'+\left({\cal H}^2+2{\cal H}'\right)\Phi=-\frac{1}{6}a^2\delta \rho_r\,.
\end{align}

Now we turn to the conservation equations. On one hand, energy conservation leads to
\begin{align}
&\delta\rho_m'+3{\cal H}\delta\rho_m+\rho_m(3\Phi'+\Delta V_m)=-a\delta Q+a\Phi Q\,,\\
&\delta\rho_r'+4{\cal H}\delta\rho_r+\frac{4}{3}\rho_r(3\Phi'+\Delta V_r)=a\delta Q-a\Phi Q\,.
\end{align}
On the other hand, momentum conservation yields
\begin{align}
&V_m'+{\cal H}V_m-\Phi=0\,,\\
&\rho_rV_r'+\frac{1}{4}\delta \rho_r-\rho_r\Phi=aQ\left(\frac{3}{4}V_m-V_r\right)\,.
\end{align}

For our purposes, it is more practical to write the above equations in terms of the isocurvature perturbation and the relative velocity respectively defined by
\begin{align}
S\equiv \frac{\delta\rho_m}{\rho_m}-\frac{3}{4}\frac{\delta\rho_r}{\rho_r}\quad{\rm and}\quad V_{\rm rel}\equiv V_m-V_r\,.
\end{align}
Using these new variables, we find that the curvature perturbation obeys
\begin{align}
\Phi''+3{\cal H}(1+c_s^2)\Phi'+({\cal H}^2(1+3c_s^2)+2{\cal H}')\Phi-c_s^2\Delta\Phi=\frac{a^2}{2}\rho_mc_s^2S\,,
\end{align}
where we defined as usual
\begin{align}
c_s^2\equiv\frac{4}{9}\frac{\rho_r}{\rho_m+4\rho_r/3}\,.
\end{align}
The isocurvature mode follows
\begin{align}\label{eq:isoeom}
S'=-\Delta V_{\rm rel}&-a(\delta Q-Q\Phi)\frac{3}{4}\frac{\rho_m+4\rho_r/3}{\rho_m\rho_r}\nonumber
\\&+\frac{2Q\rho}{a\rho_m\rho_r(\rho_m+4\rho_r/3)}\left(3{\cal H}^2\Phi+3{\cal H}\Phi'-\Delta\Phi\right)-\frac{aQ}{\rho_m\rho_r}\frac{\rho_m^2-4\rho_r^2/3}{\rho_m+4\rho_r/3}S\,,
\end{align}
and the relative velocity evolves according to
\begin{align}\label{eq:eomsvrel}
V'_{\rm rel}&+3c_s^2{\cal H}V_{\rm rel}+\frac{3}{2a^2\rho_r}c_s^2{\Delta\Phi}+\frac{3\rho_m}{4\rho_r}c_s^2S-\frac{aQ\rho}{4\rho_r}\frac{V-4V_{\rm rel}}{\rho_m+4\rho_r/3}=0\,.
\end{align}
See from Eq.~\eqref{eq:isoeom} that in the absence of energy transfer and on superhorizon scales, one has $S={\rm cnt}$.


Now we present the leading order solutions for the case of initially vanishing curvature perturbation from Ref.~\cite{Kodama:1986ud}. Deep inside the eMD, the curvature perturbation $\Phi$ is given by
\begin{align}\label{eq:conversionfinal22}
\Phi^{{\rm eISO}}_{{\rm eMD}}(k;a\gg a_{\rm eq})\approx\left\{
\begin{aligned}
&\frac{1}{5}\qquad  & k\ll k_{\rm eq}\\
&\frac{3}{4}\left(\frac{k_{\rm eq}}{k}\right)^2\qquad & k\gg k_{\rm eq}
\end{aligned}
\right.\,,
\end{align}
where the amplitude should be fixed by the initial conditions on the isocurvature perturbation. Similarly, one can study the case of an initial curvature perturbation and find that deep inside the eMD \cite{Kodama:1986ud}
\begin{align}\label{eq:curvaturefinal2}
\Phi^{e{\rm CVT}}_{e{\rm MD}}(k\gg k_{\rm eq};a\gg a_{\rm eq})\approx \frac{135}{16}\left(\frac{k_{\rm eq}}{k}\right)^4\left(\ln 4 -\frac{7}{2} +\gamma_E+\ln\left(\sqrt{\frac{2}{3}}\frac{k}{k_{\rm eq}}\right)\right)
\end{align}
where we only focused on $k\gg k_{\rm eq}$ modes since we are assuming a peaked curvature power spectrum on scales $k=k_f\gg k_{\rm eq}$, which therefore enters far inside the eRD, and we used $\gamma_E\approx0.577216$ is the Euler gamma. See how deep inside eMD the early curvature perturbation gets a suppression factor $\left({k_{\rm eq}}/{k}\right)^4$ due to decay of the modes of interest during eRD. Instead, the curvature perturbation from the conversion of early isocurvature perturbation is suppressed by $\left({k_{\rm eq}}/{k}\right)^2$. Thus, the contribution to the GW spectrum induced by the early curvature perturbation right after evaporation is suppressed by a factor $\left({k_{\rm eq}}/{k_f}\right)^8$ with respect to early isocurvature contribution. Having in mind that $k_f\gg k_{UV}$, the contribution from the early curvature perturbation right after evaporation is negligible. Note that in Ref.~\cite{Inomata:2020lmk} there is a huge enhancement of the induced GWs by the initial curvature perturbation due to the fact that the curvature power spectrum has a finite width with modes entering the horizon during eMD. In our case, the sharp peak in the curvature perturbation enters the horizon far inside eRD.

\subsection{Second order Newtonian gauge}
At second order in perturbation theory we find that tensor modes are sourced by scalar squared terms, that is
\begin{align}
h_{ij}''+2{\cal H}h'_{ij}+\Delta h_{ij}={\cal P}_{ij}\,^{ab}S_{ab}\,,
\end{align}
where ${\cal P}_{ij}\,^{ab}$ is the transverse-traceless projector defined in the following, 
\begin{eqnarray}
P_{ij}&\equiv&\delta_{ij}- \frac{\partial_i\partial_j}{\partial^2}, \hspace{22mm}A^{T}_i=P_{ij}A_j\nonumber\\
\mathcal{P}_{ijkl}&\equiv&P_{ik}P_{jl}-\frac{1}{2}P_{ij}P_{kl},\hspace{10mm}h^{TT}_{ij}=\mathcal{P}_{ijkl}h_{kl},
\end{eqnarray}
and
\begin{align}
S_{ij}&=4\partial_i\Phi\partial_j\Phi+2a^2\left(\rho_m\partial_iV_m\partial_jV_m+\frac{4}{3}\rho_r\partial_iV_r\partial_jV_r\right)
\nonumber\\
&=4\partial_i\Phi\partial_j\Phi+\frac{2a^2\rho^2}{\rho_m+4\rho_r/3}\left(\partial_iV\partial_jV
+\frac{4}{3}\frac{\rho_r\rho_m}{\rho^2}\partial_iV_{\rm rel}\partial_jV_{\rm rel}\right)
\nonumber\\
&=4\partial_i\Phi\partial_j\Phi+6c_s^2\frac{\rho}{\rho_r}\partial_i\left(\frac{\Phi'}{{\cal H}}
+\Phi\right)\partial_j\left(\frac{\Phi'}{{\cal H}}+\Phi\right)+6a^2c_s^2\rho_m\partial_iV_{\rm rel}\partial_jV_{\rm rel}\,.
\end{align}

\section{Instantaneous reheating: matching conditions\label{App:Matchingconditions}}

In this appendix we present the formulas for the matching between the eMD to the lRD used in the main text. The matching conditions are the continuity of the metric and its first derivatives in a suitable gauge as we proceed to show. 

For the background metric, the matching conditions are straightforward and requiring continuity of the metric and its first derivatives yields
\begin{align}
a(\tau<\tau_{\rm eva})=a_0\left(\frac{\tau}{\tau_0}\right)^2
\end{align}
and
\begin{align}
a(\tau>\tau_{\rm eva})=a_0\left(\frac{\tau_{\rm eva}}{\tau_0}\right)\left(\frac{2\tau-\tau_{\rm eva}}{\tau_0}\right)\,.
\end{align}
This in turns implies that the conformal Hubble parameter is given by
\begin{align}
{\cal H}(\tau<\tau_{\rm eva})=\frac{2}{\tau}\,.
\end{align}
\begin{align}
{\cal H}(\tau>\tau_{\rm eva})=\frac{2}{2\tau-\tau_{\rm eva}}\equiv\frac{1}{\bar\tau}\,.
\end{align}
This redefined conformal time $\bar\tau$ is the one we used in the main text.

The matching of first order perturbations is more subtle. From Eq.~\eqref{eq:Q} we see that in general the energy transfer due to evaporation depends on the spatial coordinates through the PBH 4-velocity. This means that in general evaporation is non-local. It only becomes local, i.e. a function of time, in the synchronous comoving coordinates \cite{Inomata:2020lmk}, where the PBH 3-velocity vanishes $V_m=0$ and the metric is given by
\begin{align}
ds^2=a^2(\tau)\left[-d\tau^2+(\delta_{ij}+2\phi_s\delta_{ij}+2\partial_i\partial_j E_s+h_{ij})dx^idx^j\right]\,.
\end{align}
In the synchronous comoving gauge evaporation takes place at the same time everywhere. Therefore, only in this gauge it is physically meaningful to require the continuity of the metric and its derivatives at evaporation. Note that the choice of the synchronous comoving gauge is possible since the PBH gas behaves as dust \cite{Malik:2008im}. 

However, the calculations of the induced GWs are considerably simpler in the Newtonian (Poisson or shear-less) gauge. Thus, for practical purposes we are interested in the matching conditions for the curvature perturbation $\Phi$ in the Newtonian gauge. After a gauge transformation, we find that the relation between the curvature perturbation $\Phi$ in the Newton and in the synchronous comoving guage is given by
\begin{align}\label{eq:Phigauge}
\Phi&=\phi_s-{\cal H}E_s'\,.
\end{align}
We see that from the continuity of the metric and its derivative in the synchronous comoving guage, we have the continuity of $\phi_s$ and $E_s$ and their derivatives. Then, Eq.~\eqref{eq:Phigauge} implies the continuity of $\Phi$ but for the continuity for $\Phi'$ requires the continuity of $E_s''$. Nevertheless, $E_s''$ is continuous as well across the transition. This is clear by looking at traceless-ij component of the Einstein equations in the synchronous gauge which yields
\begin{align}\label{eq:Es}
E_s''+2{\cal H}E_s'=\phi_s\,,
\end{align}
which tells us that the continuity of $\phi_s$ and $E_s'$ implies the continuity of $E_s''$. Thus, we have shown that the the matching conditions in the Newton gauge are the continuity of $\Phi$ and $\Phi'$.

By requiring the continuity of $\Phi$ and $\Phi'$ at evaporation, we find that $\Phi$ in lRD is given by
\begin{align}
\Phi_{lRD}(k\bar\tau)=\frac{1}{c_sk\bar\tau}\left(C_1j_1(c_sk\bar \tau)+C_2y_1(c_sk\bar \tau)\right)\,,
\end{align}
where 
\begin{align}
C_1&=-\frac{1}{8}\Phi_{MD}(c_s k\tau_{\rm eva})^{3}y_2(c_s k\tau_{\rm eva}/2)\,,\\
C_2&=\frac{1}{8}\Phi_{MD}(c_s k\tau_{\rm eva})^{3}j_2(c_s k\tau_{\rm eva}/2)\,,
\end{align}
and $j_i$ and $y_i$ are the spherical Bessel functions of order $i$. In the matching at $\tau_{\rm eva}$ we used that in a matter dominated universe $\Phi_{MD}={\rm constant}$ and $\Phi'_{MD}=0$. Since $\Phi'$ can also be written in a compact form, we also present it explicitly:
\begin{align}\label{eq:phiprime}
\Phi_{lRD}'(k\bar \tau)=-\frac{1}{\bar \tau}\left(C_1j_2(c_s k\bar \tau)+C_2y_2(c_s k\bar \tau)\right)\,.
\end{align}

\section{Induced gravitational waves integrals\label{App:kernel}}

In this appendix we derive the relevant formulas used in Sec.~\ref{Sec:IIGWS}. The Kernel of the last piece in Eq.~\eqref{eq:phsplit} has the largest contribution from the terms in Eq.~\eqref{eq:sourceterm} which contain the largest number of time derivatives. Focusing on the time derivative squared term, the Kernel of the lRD universe can be well approximated by{
\begin{align}\label{eq:kernelRD}
I_{lRD}(x,u,v,x_{\rm eva})\approx\frac{1}{2}uv\int_{x_{\rm eva}/2}^{\bar x} d{\tilde {\bar x}}\,{\tilde {\bar x}}^2\, G^{lRD}(\bar x,{\tilde {\bar x}})\frac{dT_\Phi^{lRD}(u{\tilde {\bar x}})}{d(u{\tilde {\bar x}})}\frac{dT_\Phi^{lRD}(v{\tilde {\bar x}})}{d(v{\tilde {\bar x}})}\,,
\end{align}}
where $G^{lRD}(\bar x,{\tilde {\bar x}})=\frac{{\tilde {\bar x}}}{\bar x}\sin(\bar x- {\tilde {\bar x}})$ is the Green's function during late RD and we already used switched to the variable $\bar\tau=\tau-\tau_{\rm eva}/2$ which leads to a smooth matching of the amplitude of $\Phi$ and its first derivative at time of evaporation (see App.~\ref{App:Matchingconditions} for details). From Eq.~\eqref{eq:phi}, \eqref{eq:c1c2} and \eqref{eq:xi} we have that for our scales of interest, i.e. those scales with $x_{\rm eva}\gg1$, the time derivative of the transfer function is basically given by{
\begin{align}\label{eq:tprime}
\frac{d}{d(v \tilde{\bar x})}T_{\Phi,lRD}(v\tilde{\bar x})\approx -{\cal S}_\Phi(vk)T_{\Phi,eMD}(vk)c_s\left(\frac{x_{\rm eva}}{2\tilde{\bar x}}\right)^2\sin\left[c_s v\left(\tilde{\bar x}-x_{\rm eva}/2\right)\right]\,,
\end{align}}
where we already Taylor expanded the spherical Bessel functions in the limit of large argument.

Before proceeding to the calculation of the kernel, it is important to note one we may use several approximations described in Refs.~\cite{Inomata:2019ivs,Inomata:2020lmk} to obtain an analytic expression. First, we are interested in the late time limit of the GWs evolution and we may safely send $\bar x\to\infty$. Doing this approximations to Eq.~\eqref{eq:kernelRD} and inserting Eq.~\eqref{eq:tprime} yields{
\begin{align}\label{eq:kernelRD2}
I_{lRD}(u,v,x_{\rm eva})\approx&\frac{c_s^2 uv}{32\bar x}x_{\rm eva}^4{\cal S}_\Phi(vk){\cal S}_\Phi(uk)T_{\Phi,eMD}(vk)T_{\Phi,eMD}(uk)\nonumber\\&\times\int_{0}^{\infty} \frac{dx_2}{x_2+x_{\rm eva}/2}\sin(x_1-x_2)\sin(vc_s x_2)\sin(uc_s x_2)\,.
\end{align}}
where for simplicity we have defined $x_1\equiv\bar x-x_{\rm eva}/2$ and $x_2\equiv\tilde{\bar x}-x_{\rm eva}/2$. The integral in Eq.~\eqref{eq:kernelRD2} diverges when the frequency of the faster oscillating mode equals to the sum of frequency of the slower oscillating modes, that is when $u+v=c_s^{-1}$. Such resonance is typical of systems where two modes source third one which propagates faster. This also comes apparent after integration. Fortunately, the integral with three sines in Eq.~\eqref{eq:kernelRD2} can be integrated exactly and yields
\begin{align}
\int\frac{dx_2}{x_2+x_3}&\sin(x_1-x_2)\sin(ax_2)\sin(bx_2)=\nonumber\\&
\frac{1}{4}\Bigg({\rm Ci}[(1 + a - b) (x_2 + x_3)] \sin[x_1 + (1 + a - b) x_3] \nonumber\\&+ 
   {\rm Ci}[(1 - a + b) (x_2 + x_3)] \sin[x_1 + (1 - a + b) x_3] \nonumber\\&
   - 
   {\rm Ci}[(1 - a - b) (x_2 + x_3)] \sin[x_1 - (-1 + a + b) x_3] \nonumber\\&- 
   {\rm Ci}[(1 + a + b) (x_2 + x_3)] \sin[x_1 + (1 + a + b) x_3] \nonumber\\&
   + 
   \cos[x_1 + (1 -a + b )x_3] {\rm Si}[(-1 + a - b) (x_2 + x_3)] \nonumber\\&- 
   \cos[x_1 + (1 + a - b) x_3] {\rm Si}[(1 + a - b) (x_2 + x_3)] \nonumber\\&
   - 
   \cos[x_1 - (-1 + a + b) x_3] {\rm Si}[(-1 + a + b) (x_2 + x_3)] \nonumber\\&+ 
   \cos[x_1 + (1 + a + b) x_3] {\rm Si}[(1 + a + b) (x_2 + x_3)])\Bigg)\,,
\end{align}
where $a\equiv c_s v$ and $b\equiv c_s u$. ${\rm Ci}[x]$ and ${\rm Si}[x]$ are respectively the cosine and sine integrals. For more details on this functions see Ref.~\cite{NIST:DLMF}. From this integral we can select the two most relevant contributions as follows.

We may first focus on the contribution from the resonant modes which concerns the resonant scale $a+b=1$ or $u+v=c_s^{-1}$, which is typical of the GWs induced by an adiabatic perfect fluid. Knowing that the cosine integral function diverges for vanishing argument, we select only ${\rm Ci}[(1-a-b)(x_2 + x_3)]$. Then one finds that the resonant part of the averaged kernel squared is approximately given by{
\begin{align}\label{eq:irdapp}
\overline{I^2_{\rm lRD,{\rm res}}}&(u+v\sim c_s^{-1},x_{\rm eva})\nonumber\\&\approx \frac{c_s^4u^2v^2}{2^{15}\bar x^2}x_{\rm eva}^8{\cal S}_\Phi^2(vk){\cal S}_\Phi^2(uk)T_{\Phi,eMD}^{2}(vk)T_{\Phi,eMD}^{2}(uk){\rm Ci}^2(|1-(u+v)c_s|x_{\rm eva}/2)\,.
\end{align}}

The second relevant contribution to the kernel comes from the peak in the integrand at $u\sim v\gg1$. Assuming $a=b$ (or $u=v$) and using that ${\rm Ci}(x\to\infty)\to0$ and ${\rm  Si}(x\to\infty)\to\pi/2$ we select the cosine and sine integral without $v$ dependence. In this case, the large $v$ contribution to the averaged kernel squared reads{
\begin{align}\label{eq:irdapp2}
\overline{I^2_{\rm lRD,{\rm LV}}}(u\sim v\gg 1,x_{\rm eva})\approx&\frac{c_s^4u^2v^2}{2^{13}x^2}x_{\rm eva}^8{\cal S}_\Phi^2(vk){\cal S}_\Phi^2(uk)T_{\Phi,eMD}^{2}(vk)T_{\Phi,eMD}^{2}(uk)\nonumber\\&\times\Big[{\rm Ci}^2(x_{\rm eva}/2)+\big({\rm Si}(x_{\rm eva}/2)-\pi/2\big)^2\Big]\,.
\end{align}}
Note that both kernels \eqref{eq:irdapp} and \eqref{eq:irdapp2} peak at $k=k_{\rm UV}$ as there $x_{\rm eva}$ attains the maximum value. Also notice that since we approximated the transfer function at the end of the eMD to be a broken power-law \eqref{eq:conversionfinal} with the peak at $k=k_{\rm eq}$, the GW spectrum has a contribution from the $T_\Phi^{\rm eMD}(k<k_{\rm eq})$ piece for GW momenta $0<k<2k_{\rm eq}$ and from the $T_\Phi^{\rm eMD}(k>k_{\rm eq})$ piece for GW momenta $2k_{\rm eq}>k>2k_{UV}$.

Let us turn now into the integral over momenta in the tensor power spectrum, which can be expressed as follows
\begin{align}\label{eq:prdapp}
\overline{\cal P}_{h,\rm lRD}(k)=8\int_0^\infty dv\int_{|1-v|}^{1+v}du \left[\frac{4v^2-(1+v^2-u^2)^2}{4uv}\right]^2{\cal P}_S(vk){\cal P}_S(uk)\overline{I^2_{lRD}}(u,v,x_{\rm eva})\,.
\end{align}
Dropping $k$-dependent factors we can divide the integral as follows
\begin{align}\label{eq:prdapp2}
\overline{\cal P}_{h,\rm lRD}(k)\propto&\int_0^\infty dv\int_{|1-v|}^{1+v}du {\cal P}_S(vk){\cal P}_S(uk)[...]\nonumber\\&\propto
\int_0^{v_{\rm eq}} dv\int_{|1-v|}^{{\rm min}(1+v,v_{\rm eq})}du\,v^5u^5[...]+\int_0^{v_{\rm eq}} dv\int_{{\rm max}(|1-v|,v_{\rm eq})}^{{\rm min}(1+v,v_{\rm eq})}du \,v^5u[...]\nonumber\\&
+\int_{v_{\rm eq}}^{v_{ UV}} dv\int_{|1-v|}^{{\rm min}(1+v,v_{\rm eq})}du\,vu^5[...]+\int_{v_{\rm eq}}^{v_{ UV}} dv\int_{{\rm max}(|1-v|,v_{\rm eq})}^{{\rm min}(1+v,v_{UV})}du\,vu[...]\,,
\end{align}
where $[...]$ is the remaining $u$ and $v$ dependence in the Kernel and the prefactor of the integrand in Eq.~\eqref{eq:prdapp} and we defined $v_{UV}=k_{UV}/k$ and $v_{\rm eq}=k_{\rm eq}/k$. It is easy to convince oneself that the dominant contribution for $k>2k_{\rm eq}$ is at $k\sim k_{\rm UV}$ due to the $x_{\rm eva}^8$ factor in the Kernels \eqref{eq:irdapp} and \eqref{eq:irdapp2} which originates from the large amplitude of the $\Phi'$ at evaporation. For $k<2k_{\rm eq}$ the dominant contribution to the GW spectrum is at $k=2k_{\rm eq}$ where the scalar spectrum peaks. Furthermore, since $k_{UV}\gg k_{\rm eq}$ and the integrand has a positive power of the momentum, we have that the last contribution to Eq.~\eqref{eq:prdapp2} is always the dominant part.

The integral of the momenta in the resonant case using Eq.~\eqref{eq:irdapp} is explained in detail in Sec.~\ref{Sec:IIGWS} and here we explicitly write some definitions and integrals used. We first do a change of variables given by
\begin{align}
y&\equiv \left((u+v)c_s-1\right){x_{\rm eva}}/{2}\,,\\
s&\equiv u-v\,,
\end{align}
where the Jacobian of the transformation is 
\begin{align}
|J|=\frac{1}{c_sx_{\rm eva}}\,.
\end{align}
At the end, we use that
\begin{align}\label{eq:ciint}
\int_{-\infty}^\infty dy {\rm Ci}^2(|y|)=\pi\,.
\end{align}

The integral of the momenta in the large $v$ case \eqref{eq:irdapp2} can be done as follows. We change variables to 
\begin{align}
t&\equiv u+v-1\,,\\
s&\equiv u-v\,,
\end{align}
where the Jacobian is
\begin{align}
|J|=\frac{1}{2}\,.
\end{align}
We then only focus on the $t\gg1$ contribution and we set $s=0$. This may cause some $O(1)$ uncertainty but as we later see it is not of much relevance \cite{Inomata:2019ivs}. First, for $2k_{\rm eq}>k>2k_{UV}$ we find after integrating the $t$ integral that {
\begin{align}
\overline{\cal P}_{h,lRD}(k>2k_{\rm eq})=\frac{27c_s^4}{1280\pi^2\bar x^2}&\left(\frac{3}{2}\right)^{2/3}\left(\frac{k_{\rm eva}}{k_{\rm UV}}\right)^{4/3}\left(\frac{k_{\rm eq}}{k_{\rm eva}}\right)^8\left(\frac{k}{k_{UV}}\right)^3\nonumber\\&\times\Big[{\rm Ci}^2(x_{\rm eva}/2)+\big({\rm Si}(x_{\rm eva}/2)-\pi/2\big)^2\Big]\,.
\end{align}}
We can use the large argument limit of the cosine and sine integrals to further simplify the expression. This yields an additional $(k_{\rm eva}/k)^2$ factor. Then, the spectral density of GWs reads{
\begin{align}\label{eq:omegaLV}
\Omega_{\rm GW,c,\rm LV}(2k_{\rm eq}<k<2k_{UV})=\frac{9c_s^4}{5120\pi^2\bar x^2}&\left(\frac{3}{2}\right)^{2/3}\left(\frac{k_{\rm eva}}{k_{\rm UV}}\right)^{10/3}\left(\frac{k_{\rm eq}}{k_{\rm eva}}\right)^8\left(\frac{k}{k_{UV}}\right)\,.
\end{align}}
Thus, the contribution for large $v$ is at least a factor $k_{\rm eva}/k_{UV}$ suppressed with respect to \eqref{eq:GWsres}. With a similar calculation we arrive at the expression of GW spectrum for $k<2k_{\rm eq}$, which reads {
\begin{align}\label{eq:omegaLV2}
\Omega_{\rm GW,c,\rm LV}(k<2k_{\rm eq})=\frac{c_s^4}{217500\pi^2 \,2^{2/3}3^{1/3}}\left(\frac{k_{\rm eq}}{k_{\rm eva}}\right)^{14/3}\left(\frac{k_{\rm eq}}{k_{UV}}\right)^{5}\left(\frac{k}{k_{UV}}\right)\,.
\end{align}}
If we extrapolate this results close to $k_{\rm eva}$ we see that the difference between the peak and the infrared tail is given by {
\begin{align}
\frac{\Omega_{\rm GW,c,\rm LV}(k_{\rm eva})}{\Omega_{\rm GW,c,\rm res}(k_{UV})}\approx\frac{512\,2^{2/3}}{54375\sqrt{3}\,3^{5/6}\pi}\left(\frac{k_{\rm eq}}{k_{UV}}\right)^{5/3}\left(\frac{k_{\rm eva}}{k_{UV}}\right)^2\,.
\end{align}}
Thus the spectrum at the IR tail is very much suppressed compared to the peak.

When dealing with the clustering we defined for simplicity{
\begin{align}
\tilde F_{\rm cluster}^2&\equiv{2^{1-2n}c_s^{-2n}}F_{\rm cluster}^2\nonumber\\&\times\left(2F\left[\frac{1}{2};\frac{5}{3}-n;\frac{3}{2};c_s^2\right]-\frac{4}{3}F\left[\frac{1}{2};\frac{5}{3}-n;\frac{5}{2};c_s^2\right]+\frac{2}{5}F\left[\frac{1}{2};\frac{5}{3}-n;\frac{7}{2};c_s^2\right]\right)\,,
\end{align}}
where $F\left[a;b;c;d\right]$ is the hypergeometric function.

\bibliography{biblio.bib} 

\end{document}